\documentclass[useAMS,usenatbib]{mn2e}

\usepackage{ulem}
\usepackage{amsmath}
\usepackage{amssymb}
\usepackage{graphicx}
\usepackage{natbib}
\usepackage{fixltx2e}

\newcommand\asec{^{\prime\prime}}
\newcommand\amin{^{\prime}}
\newcommand\sect{Section}
\newcommand\sectplural{Sections}
\newcommand\tabl{Table}
\newcommand\eqtn{Eq.}

\newcommand\tempbf{}
\newcommand\referee{}
\newcommand\refereetwo{}
\newcommand\refereethree{}

\newcommand {\apgt} {\ {\raise-.5ex\hbox{$\buildrel>\over\sim$}}\ }
\newcommand {\aplt} {\ {\raise-.5ex\hbox{$\buildrel<\over\sim$}}\ } 

\title[Radio halos in SZ and X-ray selected clusters]{A comparative
  study of radio halo occurrence in SZ and X-ray selected galaxy
  cluster samples}

\author[Sommer and Basu]{Martin~W.~Sommer\thanks{E-mail:
mnord@astro.uni-bonn.de (MWS); kbasu@astro.uni-bonn.de (KB)} and Kaustuv Basu
\\
Argelander-Institut f\"{u}r Astronomie, Auf dem H\"ugel 71,D-53121 Bonn, Germany}

\begin{document}

\date{Accepted 2013 October 18}

\pagerange{\pageref{firstpage}--\pageref{lastpage}} \pubyear{2002}

\maketitle

\label{firstpage}

\begin{abstract} 
  We aim at an unbiased census of the radio halo population in galaxy
  clusters and test whether current low number counts of radio halos
  have arisen from selection biases. We construct near-complete
  samples based on X-ray and Sunyaev-Zel'dovich (SZ) effect cluster
  catalogs and search for diffuse, extended (Mpc-scale) emission near
  the cluster centers by analyzing data from the NRAO Very Large Array
  Sky Survey. We remove compact sources using a matched filtering
  algorithm and model the diffuse emission using two independent
  methods. The relation between radio halo power {\refereetwo at 1.4
    GHz} and mass observables is modeled using a power law, allowing
  for a `drop-out' population of clusters hosting no radio halo
  emission. An extensive suite of simulations is used to check for
  biases in our methods. Our findings suggest {\refereetwo that} the
  fraction of targets hosting radio halos may have to be revised
  upward for clusters selected {\refereetwo using} the
  Sunyaev-Zel'dovich effect: while approximately 60\% of the X-ray
  selected targets are found to contain no extended radio emission, in
  agreement with previous findings, the corresponding fraction in the
  SZ selected samples is roughly 20\%. We propose a simple explanation
  for {\refereetwo this} selection difference based on the distinct
  time evolution of the SZ and X-ray observables during cluster
  mergers, and a bias towards relaxed, cool-core clusters in the X-ray
  selection.

\end{abstract}


\section{Introduction}

Clusters of galaxies are the relatively recent descendants of rare
high-density fluctuations in the early universe. The extensive use in
cosmology of this high-mass end of gravitationally collapsed objects
relies largely on our understanding of the properties of the
intra-cluster medium (ICM). The ICM is predominantly a fully ionized
primordial plasma containing about 90\% of the cluster baryonic mass,
and reaching very high temperatures ($T \sim 2-10$ keV) as it gathers
in the deep cluster potential well. There are two basic ways of
detecting the ICM: direct detection through the thermal bremsstrahlung
in the X-ray regime, and detection of the spectral distortion of the
cosmic microwave background radiation (CMB) in the millimeter regime
due to inverse Compton scattering. The latter is the
Sunyaev-Zel'dovich (SZ) effect
\citep{1972A&A....20..189S,1980ARA&A..18..537S}.

While surveys of galaxy clusters are concerned with the dominant
thermal component of the ICM, the plasma is also host to a population
of ultra-relativistic particles, i.e. cosmic rays. Prominent evidence
for a non-thermal population, as well as cluster-wide magnetic fields,
comes from the observation of diffuse synchrotron emission in some
galaxy clusters. These extended radio sources have a typical size of
$\sim 1$ Mpc, and are not associated with any individual
galaxies. They are broadly split into \textit{radio halos} and
\textit{radio relics} depending on their central or peripheral
position in the clusters, as well as on geometry and extent of
polarization.  While both radio relics and radio halos are thought to
be associated with cluster merger processes\footnote{ {\refereetwo see,
  e.g.},
  \citet{2001ApJ...553L..15B,2001A&A...378..408S,2004ApJ...605..695G,2010A&A...509A..68C,2013MNRAS.436..275W}
  for radio halos and
  \citet{1998A&A...332..395E,2007MNRAS.375...77H,2009A&A...494..429B,2011ApJ...735...96S,2011MNRAS.418..230V,2011JApA...32..509H,2012MNRAS.420.2006N}
  for radio relics}, radio halos are of particular interest due to
their similarity in spatial distribution with the thermal ICM
\citep[e.g.][]{2001A&A...369..441G} and their scaling with cluster
mass as measured by the $L_{\mathrm{X}}-P_{\mathrm{radio}}$
correlation \citep[e.g.][]{2000ApJ...544..686L,2009A&A...507..661B}.
As such, they are an important instrument in understanding the physics
of cluster mergers, and can possibly even trace the redshift evolution
of the cluster merger fraction.

The current understanding is that radio halos are relatively rare,
with only a few tens of objects unambiguously detected to date
\citep{2012A&ARv..20...54F}. Such low numbers stand in stark contrast
to the number of X-ray or SZ selected clusters in various all sky
surveys.  The rarity of radio halos in turn prohibits their use in the
statistical studies of large scale structure formation. However, a
possible source of bias is that almost all the current radio halo data
comes from follow-up observation of previously known X-ray clusters,
either from ad-hoc collections
\citep{2009A&A...507.1257G,2011A&A...533A..35V} or using X-ray flux
limited samples
\citep{2008A&A...484..327V,2009ApJ...697.1341R,2013A&A...557A..99K}.
Therefore, the small number of radio halos can, in part, be the result
of a selection bias that takes effect while correlating one indicator
of the thermal ICM (its total X-ray luminosity) with one indicating
its non-thermal energy. Recent discoveries of radio halos in a few
clusters with very low X-ray luminosities \citep{2011A&A...530L...5G},
as well as the lack of a prominent radio halo in some X-ray luminous
mergers \citep[e.g. A2146:][]{2011MNRAS.417L...1R} indeed raise
questions of a possible selection bias regarding radio halo
observations in X-ray selected clusters.

Predictions for radio halo counts in galaxy clusters also remain
highly uncertain, lacking a proper understanding of their origin. The
observed synchrotron emission requires acceleration of charged
particles, and there are currently two different frameworks for
mechanisms that can produce relativistic particles consistent with the
radio emission observed at GHz-frequencies. `Primary' (or turbulent
{\tempbf re-acceleration) models assume that a seed population
of high-energy electrons is re-accelerated through turbulence 
  by the second order Fermi process}
\citep{1987A&A...182...21S,2001ApJ...557..560P,2001MNRAS.320..365B},
while `secondary' (or hadronic) models rely on the continuous
injection of relativistic electrons by hadronic collisions between
thermal and cosmic ray protons
\citep[e.g.][]{1980ApJ...239L..93D,1999APh....12..169B}.  The rarity
of radio halos and a strong bi-modality of clusters in the radio/X-ray
correlation indeed suggest a preference towards the primary
(turbulent) re-acceleration models. Since the high energy protons are
long-lived, radio halos powered by electrons originating from their
collisional decay should be less sensitive to the cluster dynamical
state (but see \citet{2011A&A...527A..99E} for alterations in the
basic hadronic model to explain bi-modality). The separation of these
two models is somewhat historic, but it is unlikely that both play a
dominant role in powering radio halos. Predictions based on turbulent
re-acceleration models contain a large number of free parameters that
are matched to observations, and consequently the expected radio halo
count in the sky based on these models has considerable uncertainties
\citep{2010A&A...509A..68C}. In light of the powerful all-sky radio
surveys that are being prepared (e.g. LOFAR, ASKAP, MeerKAT) or will
become operational in the coming decade (SKA), it is urgent to
quantify the radio halo fraction in clusters in an unbiased manner.

In a previous work, we presented the first radio-SZ correlation
results for radio halos with the aim of understanding possible
selection biases and their true mass scaling \citep[][hereafter
B12]{2012MNRAS.421L.112B}. In line with the expectation that radio
halo power correlates with cluster mass, we found a clear
correspondence between these two observables. More significantly, the
strong bi-modal division present in the radio X-ray correlation
appeared much reduced. However, we could neither quantify the
selection bias nor determine the true rate of occurrence of radio
halos in clusters in a given mass bin, since the B12 results were
based on an ad-hoc selection of known radio halo clusters that were
also present in the Planck ESZ catalog
\citep{2011A&A...536A...8P}. The present work builds upon the early
results presented in B12, and constitutes the first attempt to carry
out an unbiased study of the radio halo population in an SZ selected
cluster sample.

We analyze archived data from the VLA NVSS survey
\citep{1998AJ....115.1693C} to measure the extended radio emission in
two samples of galaxy clusters and constrain the relation between mass
and radio power. Data with higher sensitivity, better $uv$ coverage
and greater resolution are available for many of our targets. However,
to avoid biasing our results towards these systems, we refrain from
using these data. We take great care to remove flux contributions from
the peripheral radio relics as well as radio lobes and other extended
emission from radio galaxies, although contamination from some radio
relics and radio mini-halos cannot be ruled out. We carry out a
two-component regression analysis to simultaneously model (i) the
scaling of radio halo power with SZ and X-ray mass observables and
(ii) the fraction of the cluster population hosting no radio halos.
One of the main scientific results of the current paper is this radio
halo ``dropout fraction" found from uniformly selected X-ray and SZ
cluster samples.

In \sect~\ref{sec:sel} we discuss the selection of our samples. 
\sect~\ref{sec:radioanalysis} is concerned with the analysis of the
radio data, and the extraction of the extended radio component from
the NVSS maps. We describe our mass-luminosity relation regression
analysis in \sect~\ref{sec:regression}, and discuss systematic effects
in \sect~\ref{sec:simul}. The results are presented in
\sect~\ref{sec:results}, {\referee along with comparisons with previous results},
and in \sect~\ref{sec:discussion} we speculate on the cause for the
measured selection difference in SZ and X-rays. We summarize our work
and present our conclusions in \sect~\ref{sec:conclusions}.  For all
results derived in this work we assume a $\Lambda$CDM concordance
cosmology with $h=0.7$, $\Omega_{m} h^2 = 0.13$ and
$\Omega_{\Lambda}=0.74$.

\section{Cluster samples}
\label{sec:sel}

Our samples are extracted from the 2013 Planck catalog of
Sunyaev-Zel'dovich sources \citep[henceforth the PSZ
sample,][]{2013arXiv1303.5089P} and a composite sample of X-ray
selected clusters extracted from the REFLEX
\citep{2004A&A...425..367B}, NORAS \citep{2000ApJS..129..435B}, MACS
\citep{2001ApJ...553..668E}, BCS \citep{1998MNRAS.301..881E} and eBCS
\citep{2000MNRAS.318..333E} clusters catalogs (henceforth the X-ray,
or simply X, sample).

The REFLEX sample covers the southern sky and is more than 90\%
complete above a flux limit of $3 \times 10^{-12}
\text{erg}~\text{s}^{-1}~\text{cm}^{-2}$ in the X-ray soft band
(0.1$-$2.4 keV). The BCS sample comprises the 201 X-ray-brightest
clusters of galaxies in the northern hemisphere with measured
redshifts $z \leq 0.3$ and fluxes higher than $4.4 \times 10^{-12}
\text{erg}~ \text{s}^{-1}~\text{cm}^{-2}$ (0.1$-$2.4 keV). eBCS is the
low-flux extension of this sample, with a corresponding flux limit of
$2.8 \times 10^{-12} \text{erg}~\text{s}^{-1}~\text{cm}^{-2}$, and
including some objects with $z > 0.3$. Although the BCS$+$eBCS sample
has 90\% completeness, it does not cover the full northern hemisphere,
and the completeness deteriorates quickly above $z > 0.3$. For these
reasons we also include the NORAS and MACS catalogs to arrive at a
nearly complete sample within our selection as described
below\footnote{See \sect~\ref{sec:results} for a discussion on how the
  completeness above $z = 0.3$ affects our results}. For all X-ray
cluster data, we use the MCXC meta-catalog of
\cite{2011A&A...534A.109P}, referring to the original catalogs only
for the uncertainties in the measured X-ray luminosities.

The selection of targets is based on considerations of (i) the
recovery of extended structures with the VLA NVSS survey, (ii) the
possible confusion of extended radio emission with radio galaxies,
(iii) the sky coverage of the NVSS survey and (iv) completeness above
a given mass threshold. The mass threshold is defined in terms of
integrated Comptonization for the PSZ sample, and in terms of X-ray
luminosity for the X-ray sample, as described below.

We consider only targets with redshifts exceeding $z = 0.1$. At this
redshift, the typical physical scale of a radio halo of one Mpc
translates into an angular scale of $\sim 9 \amin$. Although the
recovery of larger structures is in principle possible with the VLA,
we must consider that the NVSS survey is made up of snapshot
observations lacking in $uv$-coverage. As shown in section
\ref{sec:res:compare}, the NVSS derived surface brightness of targets
with confirmed radio halos approaching our thus imposed size limit
(translating into an effective redshift limit) are statistically
consistent with published results from more extensive observations,
although results for individual targets show strong variations due to
differences in the methods used for extracting the extended radio
emission.

At a redshift of 0.4, one Mpc corresponds to roughly {\referee five}
times the NVSS beam full-width at half-maximum (FWHM). At even greater
redshifts, it is conceivable that our algorithms for separating the
extended emission from point sources
(\sect~\ref{sec:radioanalysis:filter}) can fail. In addition, at high
redshifts the radio halo flux drops rapidly due to cosmological
dimming and the K-correction (\ref{sec:radioanalysis:luminosity}), so
the confusion with the other radio emitting sources in a cluster field
becomes difficult to separate in the NVSS maps.  For this reason, we
impose an upper limit of $z=0.4$.

The NVSS covers the sky above declination $-40^{\circ}$. We restrict
our samples to targets with $\delta > -39^{\circ}$ to avoid the edges
of the NVSS survey. 

As a mass measure we use $M_{500}$, the mass enclosed within the
corresponding radius $r_{500}$ inside which the mean mass density is
500 times the critical density of the universe at the redshift of the
target. In practice, a limiting mass translates into limiting values
of integrated Comptonization for the PSZ sample and X-ray luminosity
for the X-ray sample. 
In the following we discuss how masses are estimated in the two
samples, and how these are used for selecting sub-samples for the
radio halo analysis. 

\subsection{Mass estimates in the PSZ sample}
\label{sec:mass:psz}

The primary SZ observable in the PSZ sample is $Y_{500}$, the
integrated Comptonization within $r_{500}$. We denote the
\textit{intrinsic} Comptonization as
\begin{equation}
\label{eq:yszdef}
Y_{\text{SZ}} = E(z)^{-2/3} D^2_A Y_{{500}}, 
\end{equation}
where $D_A$ is the angular diameter distance and $E^2(z) =
\Omega_M(1+z)^3 + \Omega_{\Lambda} + \Omega_k(1+z)^2$ is the
normalized expansion rate of the universe. The $E(z)^{-2/3}$ term is
close to unity in the redshift range relevant for this work.

We estimate $M_{500}$ from the intrinsic Comptonization using the
baseline relation of \cite{2013arXiv1303.5080P} according to
\begin{equation}
\label{eq:szmass}
\left[\frac{Y_{\text{SZ}}}{10^{-4}~\text{Mpc}^2}\right] = 10^{-0.19}\left[ \frac{(1-b)~M_{500}}{6 \times 10^{14}~M_{\odot}} \right]^{1.79},
\end{equation}
where the factor $(1-b)$ accounts for bias due to the hydrostatic
assumption. We use the best-fit value of 0.8 for this bias
parameter. 

Because of the large beam, $Y_{500}$ cannot be measured blindly with
high accuracy from the Planck data. For this reason the PSZ catalog
offers different estimates of $Y_{{500}}$ with different size priors
from external validation \citep{2013arXiv1303.5089P}. Because X-ray
priors are not available for many clusters, we use the $Y_{z}$
estimates, which are based on breaking the size-flux degeneracy 
 with the $M_{500} - Y_{\mathrm{SZ}}$ relation
\citep[see][and references therein]{2013arXiv1303.5089P}.

\subsection{Mass estimates in the X-ray sample}
\label{sec:mass:x}

For the X-ray sample we start with the $L_{X,500} -
M_{500}^{\text{HE}}$ scaling of \citet{2010A&A...517A..92A}, using the
X-ray luminosities from the MCXC catalog
\citep{2011A&A...534A.109P}. As these masses are derived under the
assumption of hydrostatic equilibrium, we correct them by the bias
factor $(1-b)$ discussed in \sect~\ref{sec:mass:psz}. For
  brevity, we will use the notation $L_{\mathrm{X}}$ in place of $L_{X,500}$
  for the remainder of this work.

As noted by \cite{2011A&A...536A..11P}, $Y_{500}$ predicted by MCXC
X-ray luminosities are somewhat low due to the fact that X-ray
selection is more sensitive to the presence of cool-cores, and we
expect this to translate into a difference in our two mass
estimators. To test this, we directly compare the derived masses where
they are available from both the MCXC and PSZ catalogs. The result, as
expected, are slightly higher masses when using Eq.~(\ref{eq:szmass})
for the same objects, with an average ratio of $1.24$. In order to
allow similar mass cuts in both samples, we correct the MCXC masses by
this factor. We emphasize that although our mass estimators may be
biased, our main objective of having comparable mass estimates in both
samples is thus guaranteed.

\subsection{Mass selection}

We consider two types of samples, using on the one hand a
redshift-dependent mass cut, and on the other hand a constant mass
cut. We discuss these in turn.

The redshift-dependent mass selection comes from the Planck SZ cluster
sample used for cosmological analysis \citep{2013arXiv1303.5080P}, and
consists of 189 clusters of galaxies. Even though the SZ effect is
expected to provide a redshift-independent mass selection in the ideal
case, this does not account for instrumental effects. Due to the large
effective beam area of Planck, the limiting mass at which a cluster
can be reliably detected increases rapidly up to $z\sim 0.6$ and
flattens thereafter. We use this {\it COSMO} sub-sample of PSZ
clusters for our work, all but one of which have known
redshifts. After applying our redshift and declination selection we
arrive at a sample of 90 targets (henceforth the PSZ(V)
sub-sample). Translating the completeness criterion of the Planck {\it
  COSMO} clusters to the X-ray sample is difficult because the PSZ
completeness depends on the Galactic and point source masks and the
varying noise properties across the sky, the latter being a by-product
of the matched-filtering cluster detection algorithm. We therefore use
the 50\% mean completeness limit across the entire sky for our mass
threshold, as described by \cite{2013arXiv1303.5080P}, to define a
sharp (redshift dependent) mass cut for the X-ray clusters. In
conjunction with the redshift and declination cuts as above, this
yields a sample of 86 targets (henceforth the X(V) sub-sample).

Having a redshift dependent mass selection is contrary to the common
practice in previous radio halo studies, and also makes it more
difficult to address the question of what fraction of clusters host a
radio halo in a given mass bin. We therefore also consider a more
conventional redshift-independent mass cut at a constant value of
$M_{500} = 8 \times 10^{14}~M_{\odot}$. This yields 79 targets from
PSZ and 78 targets from X-ray. We denote these sub-samples PSZ(C) and
X(C), respectively. The X-ray luminosity corresponding to
  this constant mass cut is $4.6 \times 10^{44} E(z)^{7/3}$ erg/s. In
  our concordance cosmology this translates into $5.2\times 10^{44}$
  erg/s at $z=0.1$ and $7.7 \times 10^{44}$ at $z=0.4$, comparable to
  previous radio halo studies in X-ray selected clusters
  \citep[][e.g., used a constant $L_{\mathrm{X}}$ cut at $5.0 \times 10^{44}$ erg/s in
  the redshift range $0.2 < z \leq 0.4$]{2008A&A...484..327V}.  

We note that the constant-mass selected PSZ(C) sub-sample is
artificially cut because many of the cluster candidates in the PSZ
catalog lack redshifts. The high mass threshold applied in this work
nonetheless guarantees a high degree of completeness, such that a true
mass selected sample with comparable mass and redshift selection would
not make a significant statistical difference.

\begin{figure} 
  \centering
  \includegraphics[width=\columnwidth]{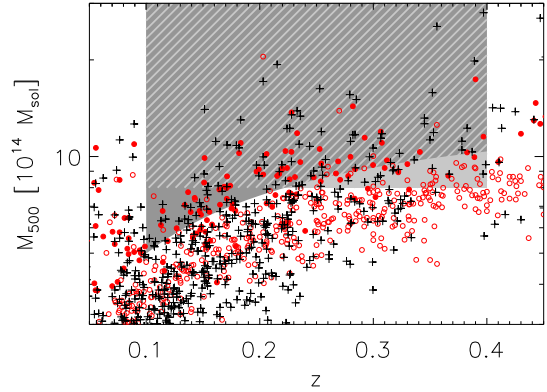}
  \caption{Mass and redshift selection. Circles represent the PSZ
    sample, with filled symbols indicating the PSZ cosmology
    sample. Plus signs represent the X-ray sample. Our two
    selection functions are indicated by the shaded regions: {\tempbf a
    constant mass threshold of $8\times 10^{14}M_{\odot}$ in light gray, 
    and a redshift dependent mass threshold defined to mimic the PSZ
    cosmology sample selection in dark grey (see text). The hatched region marks the 
    overlap between the two selections.}}
  \label{fig:selection}
\end{figure}

\begin{figure} 
  \centering
  \includegraphics[width=\columnwidth]{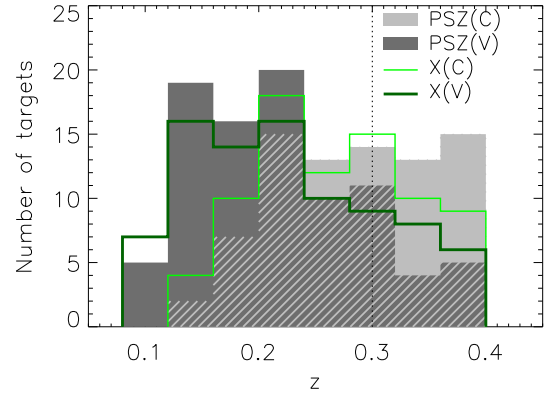}
  \caption{Redshift distribution in the two samples and using the two
    different mass selection criteria ({\referee the hatched region is the overlap of the PSZ(V) and PSZ(C) samples}). {\referee The redshift
      distributions of the PSZ and X samples start to diverge slightly
      at approximately $z=0.3$, indicated by the vertical dotted
      line.}}
  \label{fig:zhists}
\end{figure}

As part of the analysis, we remove NVSS fields in which bright point
sources and their side lobes cause our point source removal algorithm
(\sect~\ref{sec:radioanalysis:filter}) to fail (in the sense that it
does not converge on a point source free map). 
The mass and redshift selections are summarized in Table \ref{tab:sample} and 
Fig.~\ref{fig:selection}. The redshift distribution of the selected
sources are similar between the two main samples for each type of mass
selection, as is illustrated in Fig.~\ref{fig:zhists}.

\begin{table}
  \caption{Cluster samples. The primary selection is based on the redshift, declination, and the $Y_{\text{SZ}}$ (or $L_{\mathrm{X}}$) cut. Subsequent data quality control removes objects with strong interferometric noise artifacts around bright ($\apgt 1$ Jy) 
    anywhere in the field. The designators V and C indicate the redshift dependent and constant mass selections, respectively.}       
\label{tab:sample}      
\centering                 
\begin{tabular}{l c c c c}
\hline\hline               
Sub-sample & Mass & Primary    & Flagged due & Final \\
          & limit & selection & to bad data       & sample \\
\hline\hline
PSZ(V)           & $z$-dependent & 90 & 1 & 89 \\
X(V) & $z$-dependent & 86 & 1 & 85 \\
\hline
PSZ(C)           & $8\times 10^{14}M_{\odot}$ & 79 & 0 & 79 \\
X(C) & $8\times 10^{14}M_{\odot}$ & 78 & 1 & 77 \\
\hline
\end{tabular}
\end{table}

\section{Analysis of the NVSS data}
\label{sec:radioanalysis}

For each field in our two samples, we obtain a $1.3^{\circ}\times
1.3^{\circ}$ 1.4 GHz radio map from the NVSS archive. The NVSS
synthesized beam has a FWHM of 45$\asec$, and the rms noise is about
0.45 mJy/beam. These numbers are approximately uniform across the
entire survey area.

In \sect~\ref{sec:radioanalysis:filter} we describe the different
filtering steps applied to the radio maps in order to separate the
extended radio emission from other components such as background and
foreground point sources and radio galaxies (typically brightest
cluster galaxies, BCG). For simplicity, we refer to these
contaminating sources simply as radio point sources, although in many
cases they may not be point-like with respect to the NVSS beam. In
\sect~\ref{sec:radioanalysis:fluxextraction} we discuss two
independent methods of measuring the extended radio emission.

\subsection{Filtering of the maps}
\label{sec:radioanalysis:filter}

Because point sources cover a wide dynamic range in flux density and
considering the relatively poor resolution of NVSS, there is no single
method that will reliably separate them from an extended component. We
have considered and tested several methods of extracting the extended
emission, and have found the direct removal of point sources to be the
most robust. A matched filter applied directly to the extended
emission is found to work poorly because the tails of compact sources
are found to contaminate the extracted signal. In addition, the
morphology of the sought radio emission is not known \textit{a
  priori}.

The multi-scale spatial filter of \cite{2002PASP..114..427R} was used
for similar analyses by, e.g., \cite{2009ApJ...697.1341R} and
\cite{2011ApJ...740L..28B}. This filtering method can cause a low-flux
bias in case there are significant substructures on the diffuse
emission, and can potentially suffer from contamination due to the
tails of bright sources. As there is furthermore no clear definition
of the recovered scale, we will not use this filtering method.

We follow three separate, consecutive steps for the removal of point
sources. Regions around bright sources ($S_{1.4} \apgt 20$ mJy) are
completely cut from the map, weaker sources are fitted with Gaussians
and subtracted from the map, and finally the residual map is low-pass
filtered to remove point source emission below the detection
threshold.

We identify point sources by a matched filtering algorithm. Because
compact radio sources can be slightly resolved by the NVSS beam, we
pre-smooth the NVSS map with a Gaussian in order to allow for a better
matching of sources to a point source template. The level of smoothing
is based on the maximum size of a radio galaxy belonging to the
cluster, which we assume to be 100 kpc. The angular equivalent of this
extent at the redshift of the cluster, $\Theta_{100\,\text{kpc}}$, is
used as the FWHM of a Gaussian with which we smooth the map. The final
resolution of the map is thus
\begin{equation}
\Theta_{S} = \sqrt{\Theta_{100\,\text{kpc}}^2 + \Theta_{\text{NVSS}}^2},
\end{equation}
where $\Theta_{\text{NVSS}}$ is the 45$\asec$ FWHM of the NVSS
beam. The intrinsic NVSS resolution is effectively degraded by 8\% at
$z=0.4$ and by roughly 60\% at $z=0.1$.  We construct a template
source with FWHM $\Theta_{S}$ as input for the matched filter. In
order to avoid contamination from the extended radio emission, we
taper scales larger than three times the NVSS beam FWHM in the matched
filtering.

The matched filtering algorithm is iterative in the sense that we
re-estimate the noise power spectrum in the 1.4 GHz radio map after
removing each point source, starting with the brightest one, in order
to successively increase the accuracy of the filter. Assuming that
there are no preferred directions in the map, the noise power spectrum
is radially averaged.

The peak of the brightest radio source in the filtered map can be
offset for two reasons: the radio source can be extended with respect
to the map resolution, and the noise power spectrum can be
overestimated. In fact, in most cases both effects are strong enough
to affect the result. Thus, in each iteration we fit an elliptical
Gaussian model to the radio source in the original unfiltered map. For
the modeling we constrain the centroid position to be within
$\frac{1}{5}$ of the extent of the NVSS beam FWHM. We also constrain
the \textit{intrinsic} (deconvolved) source FWHM to be less than 100
kpc (major and minor axes). 

Sources brighter than 20 mJy are found to leave residual structures
after being subtracted, due to the NVSS beam not being a perfect
Gaussian. When such a source is found by our algorithm, we flag a
region around the source corresponding to a radius where the model
flux has dropped below 0.1 mJy/beam, which is well below the rms level
of the NVSS survey. For any source below 20 mJy, the model is
subtracted from the map before a new noise power spectrum is
estimated and the next iteration takes place.

\begin{figure} 
  \centering
  \includegraphics[width=\columnwidth]{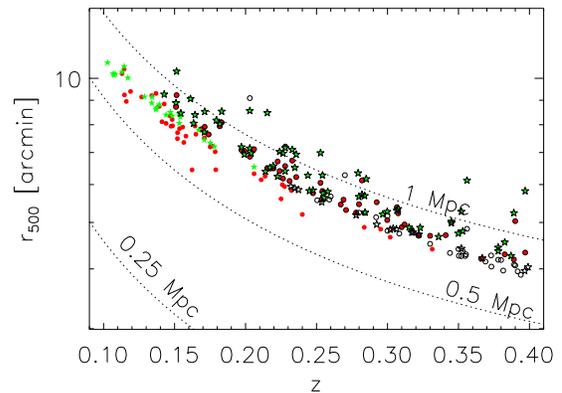}
  \caption{Physical scale $r_{500}$ vs. redshift for the PSZ (circles)
    and X-ray (stars) samples. Filled symbols indicate the redshift
    dependent (V) samples while open symbols represent the samples
    selected by constant mass (C). Dotted lines indicate constant
    physical scales.  }
  \label{fig:zr500}
\end{figure}

In each iteration we also re-estimate the rms of the radio map, and
stop iterating at a radio source peak signal-to-noise ratio of 3. At
this significance level a relatively high number of spurious
detections is expected; thus we take care to model both positive and
negative peaks. Conversely, there will be a significant population of
sources below our significance threshold, which can seriously affect
the subsequent measurement of the extended component. For this reason
we use a Butterworth filter to low-pass filter the residual map,
effectively removing remaining structures smaller than 200 kpc in
extent. Compared to $r_{500}$, this scale is very small
(Fig.~\ref{fig:zr500}). Consequently, this low-pass filter will not
affect the sought-after radio halo emission in a significant way, but
will minimize the contamination from radio mini-halos which are
comparable to the chosen filter size.

\subsection{Extraction of the extended signal}
\label{sec:radioanalysis:fluxextraction}

Measuring the luminosity of the extended radio emission component in
any single galaxy cluster using the relatively shallow NVSS data is
the central challenge of this work. We consider two methods
here. First, we take the approach of fitting a model to each
target. In most cases the signal-to-noise ratio of a single target
will be too low to allow for a detailed modeling of the emission. For
this reason, we make the simplifying assumption of self-similarity,
allowing us to stack our samples to obtain a high-fidelity
measurement of the mean radial profile. This approach is considered in
\sectplural~\ref{sec:radioanalysis:profile} and
\ref{sec:radioanalysis:luminosity}

Because the assumption of self-similarity may not be valid, we
consider a second approach in which we integrate the flux in an
appropriate aperture (we will use $0.5 \times r_{500}$ for reasons
explained below). This method is complicated by the fact that because
we have blanked out several regions of most radio images, often
including bright radio galaxies in the centers of galaxy clusters,
there is missing information which in many cases cannot be reliably
recovered. We consider the method in detail in
\sect~\ref{sec:radioanalysis:directint}.

\subsubsection{Mean profile of the radio halo emission}
\label{sec:radioanalysis:profile}

In order to stack the maps, it is necessary to first scale them both
to a common linear scale and to a common normalization to
avoid over-weighting the radio-bright targets. We discuss each of
these aspects in turn.

Scaling to a common linear scale is difficult because little is known
about the expected physical extent of the radio
emission. \cite{2007MNRAS.378.1565C} constrained the linear sizes of
radio halos by averaging their minimum and maximum extensions, and
found a strong correlation of this size measure, $R_H$, with radio
halo power $P_{1.4 \, \text{GHz}}$. Although $R_{H}$ would seem a
useful measure for the size of a radio halo we note that the scaling
of $r_{500}$ and $R_H$ is non-linear, for which we cannot provide a
feasible physical explanation. In this work, we scale the radio
emission using $r_{500}$. This is a robust physical scale of clusters
of galaxies and should also have a physical meaning for radio halos
considering the strong correlation between radio halo power and SZ
signal in a similar aperture shown in B12.

Scaling the normalization would ideally be done by dividing each map
by the radio flux at some common physical radius. However, again due
to the limited significance that we expect for most of our
measurements, this is not possible. Instead we choose to scale each
map by the total radio power. This poses the additional problem that
the total radio power is not known until we have chosen the
appropriate model to fit. It is thus necessary to proceed in an
iterative way, where we first derive a radial profile (with the model
described below) from a non-normalized stack (possibly biased towards
radio-bright targets), then fit for total power, and finally derive a
new profile from a stack where each map has been scaled according to
the radio power thus found. We find that this approach converges very
quickly, and that only two iterations are necessary.

In the first iteration, we assign equal weights to all fields, since
the NVSS survey has very close to uniform rms. We then re-scale all
radio images to $r_{500}$, adjusting the weights accordingly, before
stacking the images and constructing the mean radial profile.

While we find that a number of models fit the resulting profiles,
simple models with only one `shape' parameter (such as a Gaussian) are
insufficient. For reasons of convenience, we perform a fit using the
functional form of the isothermal $\beta$-model: 
\begin{equation}
  S = S_0 \left( 1 + \left( \frac{x}{x_s} \right) ^{2} \right) ^{(1-3\beta)/2}.
\end{equation}
Here $S_0$ is an arbitrary normalization of the amplitude, and $x
\equiv r / r_{500}$. The shape of the profile is determined by the
parameters $x_s$, the scale radius in units of $r_{500}$, and $\beta$,
which relates to the outer profile slope. Note that we use the
$\beta$-model simply as a flexible parametric model to fit the
extended emission $-$ there is no attempt at establishing a formal
connection between the radio halo and the cluster SZ profile.

The fit is performed inside $0.5 \times r_{500}$, corresponding to a
median angular radius of $\sim 2.8 \amin$ in the PSZ sample, and a
similar radius in the X-ray sample.  We restrict the fit inside $0.5
\times r_{500}$ for two reasons: (i) we want to minimize the
contamination from extended radio relics that are typically found in
the cluster outskirts, and (ii) there can be some attenuation of the
radio emission in the outskirts due to the limited $uv$ coverage in
the NVSS maps.

\begin{figure} 
  \centering
  \includegraphics[width=\columnwidth,trim=0 0 0 25,clip=true]{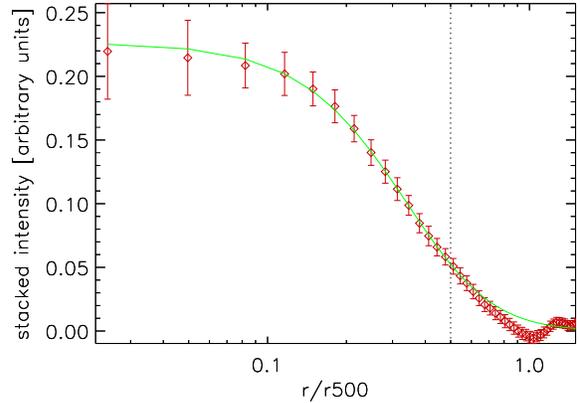}
  \caption{Stacked radial profile of radio emission normalized to
    $r_{500}$ and total radio power for the PSZ sample.  The best-fit
    model is indicated by the solid curve. Only data within $0.5
    \times r_{500}$ (dashed line) is used for the fit.}
  \label{fig:profiles}
\end{figure}

We carry out the radial fits separately for each sample, and find
consistent results from the X-ray and PSZ samples. {\referee Combining in quadrature}, we use the derived parameters
to obtain total radio power measurements as discussed in
\sect~\ref{sec:radioanalysis:luminosity}, and use the results to
re-normalize our maps for a new stacked profile, the result of which
is shown for the PSZ sample in Fig.~\ref{fig:profiles}.

The results of the second fit are presented in
\tabl~\ref{tab:betafit}. We verify that an additional iteration
produces results consistent with these. As the results from the two
samples are consistent within uncertainties, we combine them to obtain
a `universal' profile, scaled at $r_{500}$.  We use this profile to
constrain the radio luminosity of each target as described next in 
section \ref{sec:radioanalysis:luminosity}.

\begin{table} 
\caption{Stacked radial fit results}       
\label{tab:betafit}      
\centering                 
\begin{tabular}{l c c}
\hline\hline               
Sample & $x_s$ ($r_{500}$) & $\beta$ \\ 
\hline
   PSZ  & 0.479 $\pm$ 0.105 & 1.673 $\pm$ 0.416 \\      
   X-ray  & 0.468 $\pm$ 0.062 & 1.850 $\pm$ 0.285 \\
\hline
   combined & 0.471 $\pm$ 0.048 & 1.793 $\pm$ 0.235 \\                  
\hline                                   
\end{tabular}
\end{table}

To investigate whether there is a mass dependence in the radio surface
brightness profile, we also divide each sample into two bins, dividing
at the median Comptonization of the PSZ sample and the corresponding
X-ray luminosity. Within uncertainties we find no evidence of a mass
dependence in the best-fit profile, and thus proceed with a common
radial profile for all objects. 

\subsubsection{Radio measurements by radial fitting}
\label{sec:radioanalysis:luminosity}

In each cluster field, we use the radial model derived in the previous
subsection, with the parameters $x_s$ and $\beta$ fixed to their
combined best-fit values (bottom row of \tabl~\ref{tab:betafit}), to
fit for the peak amplitude of the radio emission. 
  Integrating over the resulting model out to $0.5 \times r_{500}$, we
  derive the total radio power (within $r < 0.5 \times r_{500}$) from
  the flux $S_{\mathrm{1.4 \, GHz}}$ according to 
\begin{equation}
P_{\mathrm{1.4 \, GHz}} \ = \ (4\pi \ D_L^2) \ S_{\mathrm{1.4 \, GHz}} 
\ \frac{ {\cal K}(z) } { (1+z) },
\label{eq:lumeq}
\end{equation} 
where $S_{\mathrm{1.4 \, GHz}}$ is the integrated flux density, $D_L$
is the cosmological luminosity distance and ${\cal K}(z)$ is the
K-correction, accounting for the fact that due to the redshift, the
observed flux corresponds to a rest frequency higher than 1.4 GHz. To
compute the K-correction term, we assume that the radio flux scales
with frequency as $S_{\nu} \propto
\nu^{-1.2}$. \citet{2012A&ARv..20...54F} report spectral slope
measurements over tens of radio halos, where the median is consistent
with our assumed slope.

Integrating the radial profile only to half of $r_{500}$ implies that
a significant fraction of the total radio flux is unaccounted
for. Assuming the radial profile derived in \sect~
\ref{sec:radioanalysis:profile}, we account for this loss. Integrating
the profile over the azimuthal angle and in radius out to $0.5 \times
r_{500}$ and to $r_{500}$ results in an effective correction factor of
{\referee 1.46}.  Applying this correction, our quoted
radio measurements are scaled to the region within $r_{500}$. {\referee
  We rely on our model rather than the data for this extrapolation
  because NVSS maps may not be sensitive to the largest scales (indeed
  some maps show negative artifacts at large radii), while quoting
  values only within $0.5 \times r_{500}$ will cause a systematic
  under-estimation of the total radio halo fluxes when comparing with
  published results (see \sect~\ref{sec:indiv:halo}). The results
  of our work do not depend on this extrapolation.}

The dominant source of statistical uncertainty in
$P_{\mathrm{1.4~GHz}}$ is the residual noise due to unfiltered point
sources and other compact objects in the map.  We estimate this
uncertainty in the flux measurement by inserting each fitted profile
into 100 noise realizations based on the noise power spectrum of the
filtered radio image and re-performing the fit. In order for the data
to carry the same weight, we mask out the same regions in each mock
image as were masked in the real radio map due to contamination by
bright sources (see \sect~\ref{sec:radioanalysis:filter}). 
We verify that the mean flux found by this method is consistent with
the nominal flux in the map, and use the standard deviation of the
integrated flux from the 100 realizations as an estimate of the
uncertainty in our radio flux.

\subsubsection{Radio measurements by direct integration}
\label{sec:radioanalysis:directint}

To verify that our flux extraction method is robust, we perform a
separate analysis based on extracting the diffuse radio flux by direct
integration in the map rather than by fitting a model. Even though the
radial fitting method takes into account a variation in radio halo
sizes scaled to the cluster mass, actual scaling of radio halo radii
can be different, affecting our flux measurements. 

The method of direct summation of flux inside the core region of
clusters is somewhat complicated by the fact that the processed NVSS
maps may have ``holes'' at the positions of bright point sources
(\sect~\ref{sec:radioanalysis:filter}). We fill in the holes by
interpolating from the unmasked pixels at the edges, using the inverse
distance squared between the pixels as the weight function.  We note
that this process will tend to underestimate the flux in cases where
the peak of the diffuse radio emission coincides with a strong point
source; however, simulations with radio emission constructed from the
profile found in the previous subsection and with bright point sources
added in the center indicate that the bias is at a level of 5\% for
the high redshift clusters, and much less at the median redshifts of
our samples.  We measure the extended radio flux in each NVSS map by
summing the flux in an aperture, again corresponding to $0.5 \times
r_{500}$, and proceed to compute the luminosity as described in the
previous subsection.

\section{Regression analysis}
\label{sec:regression}

In this section we outline our procedure for finding the best-fit
empirical relation between radio power and the mass observable
($Y_{\text{SZ}}$ in the case of the PSZ sample and $L_{\mathrm{X}}$ in the
case of the X-ray sample). We model the dependence using a
power law, and perform the regression taking into account
uncertainties in both coordinates, intrinsic scatter and what we call
a {\it dropout fraction} term, quantifying the fraction of data points
consistent with not belonging to the main distribution.

Regression of scaling relations is traditionally carried out in
log-log space, where the problem of fitting a power law reduces to a
linear regression problem. Even when errors are not log-normally
distributed (which is rarely the case), this method can provide a good
approximation of the assumed underlying power law, provided that
measurement uncertainties are small. However, due to our data having
relatively large uncertainties, particularly in the total radio power,
and the measurements following close to normal distributions, we fit
the scaling relations to the measurements directly in linear
space. This also has the advantage that no special provisions are
needed for non-detections; we can use all measured values (also
negative ones) along with their uncertainties in a uniform way.

We fit the data to a power law relation of the form
\begin{equation}
  \label{eq:powerlaw:general}
  y = f(x) = A x^{B},
\end{equation}
where, in our case, $y$ is the total radio power, and $x$ is the mass
observable. We include in our analysis a fractional intrinsic scatter
$\sigma_{\mathrm{F}}$ in radio power, such that the linear scatter at
a given power is equal to $y \sigma_{\mathrm{F}}$. Note that just as
fitting in log-log space implies log-normal scatter when using a
$\chi^2$ formalism, fitting in linear space likewise implies some form
of linear scatter (since scatter and uncertainties are on a equal
footing in this formalism). Although this has the unfortunate
disadvantage of allowing the $y$ variable to scatter below zero, which
is unphysical, we find that a fractional scatter provides the best fit
to our data.

There is a plethora of $\chi^2$-based methods described in the
literature for finding the best-fit model when uncertainties are
present both the dependent and independent variables. Here we follow a
maximum-likelihood approach similar to that described by
\cite{2007ApJ...665.1489K}, and we refer to that work for a detailed
discussion. It is assumed that the independent variable is drawn from
a distribution which is approximated by a mixture of $K$ Gaussians
with normalization $\pi_k$, mean $\mu_k$ and variance
$\tau^2_k$. Denoting the measured data as $\mathbf{z} =
(\mathbf{y},\mathbf{x})$, the measured data likelihood function for
data point $i$ is then a mixture of normal distributions with weights
$\pi_k$, means $\zeta_k = (f(\mu_k),\mu_k)$ and covariance matrices
$\mathbf{V}_{k,i}$. The likelihood function is given by the product of
the individual likelihoods of the data points,
\begin{equation}
{\cal L} = \prod_{i=1}^{n}\sum_{k=1}^{K}\frac{\pi_k}{2 \pi |\mathbf{V}_{k,i}|^{1/2}} \times e^{-\frac{1}{2} (z_i - \zeta_k)^\mathrm{T} \mathbf{V}^{-1}_{k,i} (z_i - \zeta_k)}.
\label{eq:l}
\end{equation}
{\referee We again refer to
  \citet[][{\refereetwo their section 4.1}]{2007ApJ...665.1489K} for a detailed discussion on how the
  distribution of the independent variable is realized in practice.} 
The covariance matrix of the $i$th data point and the $k$th Gaussian
is given by
\begin{equation}
  \textbf{V}_{k,i}  =  \left( \begin{array}{cc}
      \beta_i^2 \tau_k^2 + \sigma^2 + \sigma^2_{y,i} & \beta_i \tau_k^2 + \sigma_{xy,i} \\
      \beta_i \tau_k^2 + \sigma_{xy,i} & \tau_k^2 + \sigma^2_{x,i} \end{array} \right),
\end{equation}
where in our case $\sigma = y\sigma_{\mathrm{F}}$ and $\beta_i
=\frac{\partial f}{\partial x}|_{x_i}$. {\referee The covariances
  $\sigma_{xy,i}$ are all assumed to be zero, while the variances
  $\sigma^2_{x,i}$ and $\sigma^2_{y,i}$ are derived from the
  individual measurement uncertainties on the mass estimators and
  radio halo powers, respectively.}

There are several ways to accommodate the possibility of a bi-modality
in $f(x)$. The most simple method would be to carry out the analysis
excluding any data points that by some selection criterion are deemed
to be consistent with zero radio luminosity. This can be accomplished,
for example, by rejecting all data points below some significance
threshold. However, such an algorithm is likely to introduce a bias
in the estimated parameters because data points at the low $x$ end are
also expected to have less radio emission.

A more robust estimate can be achieved by introducing a measure of
bi-modality in the likelihood computation. To this end, we introduce a
set of $N$ binary parameters $q_i$ in the likelihood, each of which is
unity if the corresponding data point belongs to the ``on-population''
(the sought power law) and zero if the corresponding data point
belongs to the ``off-population'' (no extended radio emission). The
individual likelihood of data point $i$ needs to be modified in the
case that $q_i=0$, in the sense that $\zeta^{\mathrm{off}}_k = (0,\mu_k)$ and the
covariance matrix $\textbf{V}_{k,i}$ reduces to the simple form
\begin{equation}
  \textbf{V}^{\mathrm{off}}_{k,i}  =  \left( \begin{array}{cc}
      \sigma^2_{y,i} & \sigma_{xy,i} \\
      \sigma_{xy,i} & \tau_k^2 + \sigma^2_{x,i} \end{array} \right)
\end{equation}
under the (rather straightforward) assumption that the off-population
has zero mean and no intrinsic scatter. Including the new-found
parameters in our analysis, we can still use Equation~\ref{eq:l} to
compute the likelihood. 

The inclusion of the $q_i$ implies that we now have more parameters
than data, this is not generally a reason for panic, in particular
because each $q_i$ can only take on two discrete values. More
importantly, however, we can marginalize over the nuisance parameters
$q_i$ by replacing them with a single continuous variable, the
\textit{dropout fraction}, $g$, signifying the fraction of
off-population measurements \citep[see, e.g.,][for a similar
argument]{2010arXiv1008.4686H}. The likelihood for an individual data
point then becomes
\begin{equation}
\begin{split}
{\cal L}_{k,i} =  & \frac{(1-g)\,\pi_k}{2 \pi |\mathbf{V}_{k,i}|^{1/2}} 
e^{-\frac{1}{2} (z_i - \zeta_k)^\mathrm{T} \mathbf{V}^{-1}_{k,i} (z_i - \zeta_k)}  \\
 & + \frac{g \, \pi_k}{2 \pi |\mathbf{V}^{\mathrm{off}}_{k,i}|^{1/2}} 
e^{-\frac{1}{2} (z_i - \zeta_k^{\mathrm{off}})^\mathrm{T} (\mathbf{V}^{\mathrm{off}}_{k,i})^{-1} (z_i - \zeta_k^{\mathrm{off}})}.
\end{split}
\end{equation}
This so-called \textit{mixture model}\footnote{``Mixture'' here refers
  to the mixture of two populations in the data, and should not be
  confused with the mixture of Gaussian functions for the distribution
  of the independent variable.} is used for the remainder of this work.

We use a Bayesian method to estimate the best-fit parameters of the
power law through a Markov Chain Monte Carlo (MCMC) approach.  We run
four parallel chains starting from dispersed initial values to check
for convergence by correlating the resulting posterior distributions.
The $68\%$ confidence interval is computed by integrating down from
the maximum likelihood peak. We note that the dropout fraction and the
intrinsic scatter have sharp boundaries at zero, imposed by physically
motivated priors.

\section{Control of systematics}
\label{sec:simul}

Our final result will be the product of a rather complex analysis,
involving filtering and flagging of the raw NVSS images as well as a
non-standard method for estimating the power law relating the mass
observable to radio power under the assumption of bi-modality. In this
section we describe a series of simulations aimed at taking into
account effects of radio point sources below the detection threshold,
the flux extraction methods, and the fitting procedure. We rule out
the possibility of bias due to radio lobes, and perform several null
tests to check for residual source contamination or ``clean bias" in
the NVSS maps. { \referee In addition, we compare individual radio halo
  measurements to published results where available.}

\subsection{Model fitting}
\label{sec:sim:fit}

To test the model fitting analysis, we perform a Monte-Carlo
simulation taking the best-fit parameter values from each sample ({\referee as
obtained from the regression analysis of our X-ray and SZ samples; see}
\sect~\ref{sec:results}) as input, and verifying that the results of
the regression analysis are consistent with these input parameters.

We use the measured values of the independent variable
($Y_{\text{SZ}}$ or $L_{\mathrm{X}}$) as a starting point, and derive
theoretical values of the dependent variable (radio power) using our
bi-modal distribution: For each data point we use the best-fit power
law parameters (with probability $1-g$) or set the radio halo power to
zero (with probability $g$). Random noise at the level of the actual
measurements is added to both variables, as well as random scatter in
the dependent variable. This procedure is repeated several thousand
times, and in each iteration we derive the best-fit parameters from
the simulated data points.

\begin{figure} 
  \centering
  \includegraphics[width=\columnwidth]{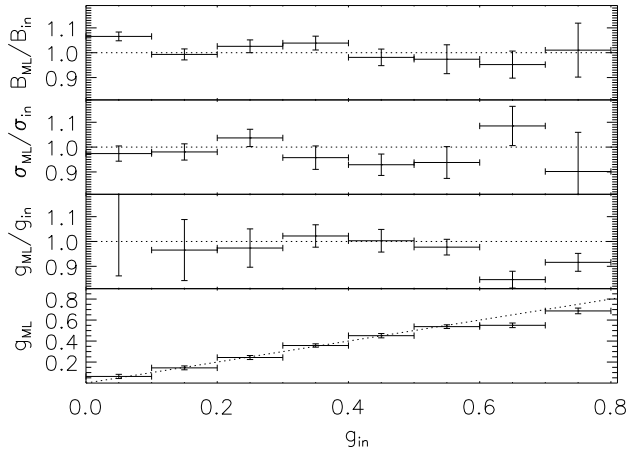}
  \caption{Recovery of model parameters (power law slope $B$, linear
    scatter $\sigma$ and dropout fraction $g$) as a function of the
    input dropout fraction $g_{\mathrm{in}}$ from a set of 3000
    PSZ-like simulations. The dotted lines indicate the input
    parameter values, which were kept constant excepting the dropout
    fraction. Vertical error bars indicate the spread in fitted
    (maximum likelihood) parameter values, and horizontal error bars
    indicate bin widths. Subscripts `ML' and `in' indicate maximum
    likelihood and input values, respectively. {\refereetwo The
      seemingly peculiar behavior at $g_{\mathrm{in}} = 0.65$ is the
      result of a statistical fluctuation slightly in excess of one
      sigma.}}
  \label{fig:fitbias}
\end{figure}

We also allow the dropout fraction $g$ to vary in our simulation to
investigate whether it can be biased low by our fitting method. In
general, we find this not to be the case, as illustrated in
Fig.~\ref{fig:fitbias} for the PSZ sample.  The X-ray sample yields
similar results. We also find that the input parameters, including the
scatter, can be recovered to an accuracy better than the statistical
uncertainties (Fig.~\ref{fig:fitbias}).
We find a slight
underestimate of the intrinsic scatter, at a level of up to 5\% in the
PSZ sample and up to 7\% in the X-ray sample, in the range of $g$
consistent with the respective best fits. We note, however, that this
bias is well below the level of the statistical uncertainties in this
parameter.

\subsection{Effects of Filtering}
\label{sec:sim:filter}

We now turn our attention to any possible bias arising from the
filtering of the radio maps. To this end, we fabricate $10^4$ maps
with simulated extended radio signals using the radial profiles
derived in \sect~\ref{sec:radioanalysis:profile}. We generate each
radio model randomly with model parameters in the range $0.1 < z <
0.4$, $0.8~\text{Mpc} < r_{500} < 1.5~\text{Mpc}$ and $0 < \log_{10}(S
[\text{mJy}]) < 2.3$, resulting in a distribution that is by no means
meant to be realistic, but will serve our purpose of investigating the
effects of filtering. We add each radial model to a randomly chosen
patch from the NVSS survey to ensure realistic noise properties. We do
not add a population of radio point sources at this point, but
consider this problem separately in \sect~\ref{sec:sim:ptsrc}

After passing the simulated maps through our filtering apparatus, we
consider the fraction of the input flux that is recovered after
filtering and point source blanking. Using both the radial fit and
direct integration methods, we find that the recovered flux is very
close to 100\% of the input flux, with no detectable dependence on the
level of the input flux or the linear size of the emission. 

\subsection{Effects of a faint radio point source population}
\label{sec:sim:ptsrc}

It is possible that a population of faint point sources below the
detection limit can mimic a diffuse extended radio
emission. Integrating the 1.4 GHz volume-averaged cluster luminosity
function of \cite{2004ApJ...617..879L} for luminosities below that
corresponding to the completeness limit of the NVSS survey at the
median redshifts of our samples results in a total point source
contribution of several mJy assuming typical values of $r_{500}$,
which is similar to typical flux levels of detected radio emission at
the low-mass end of our samples. We therefore carry out a simulation
to investigate whether there is a residual population of faint point
sources that can contaminate our signal after the filtering process.

We model the cluster radio point source population from the luminosity
function and radial distribution of the AGN and star-forming (SF)
galaxies. For the AGN population, we use the luminosity function and
radial distribution of \cite{2011A&A...529A.124S}, who determined the
volume averaged radio luminosity function at 1.4 GHz in cluster
environments from an optical sample of clusters and groups, to
populate simulated fields with sources. For the star-forming galaxies,
we use the luminosity function of \cite{2004ApJ...617..879L}, which
was determined at lower redshifts than that of Sommer et al. and thus
has leverage on this relatively fainter population. We consider
luminosities down to a limit of $10^{19}$ W Hz$^{-1}$. Extrapolating
the luminosity function below this limit does not lead to an
appreciable increase in total luminosity. The radial distribution of
star-forming galaxies is difficult to decouple from that of the AGN
population\footnote{The opposite is not the case: at high redshifts,
  point source counts above the completeness limits of the NVSS and
  FIRST surveys are completely dominated by AGN.}. Here we
differentiate the integrated counts given by
\cite{2011MNRAS.416..680C} to arrive at a radial profile,
 to populate simulated fields with the
distribution of star-forming galaxies given by the luminosity
function.

In simulating cluster fields, we follow the same procedure as outlined
in \sect~\ref{sec:sim:filter}, the only difference being the inclusion
of the cluster point source populations. We also generate three sets
of reference maps, containing (i) only the extended signal, (ii) only
the point sources, and (iii) neither of these components.

After filtering, we investigate whether the output maps contain more
signal than the reference maps with no point sources. We find this
indeed to be the case, although the effect is small: For a halo with
$r_{500} = 1~\text{Mpc}$, the residual power is on the order of $1
\times 10^{23}$~W Hz$^{-1}$, corresponding to less than $1$~mJy at
$z=0.25$.

We find that, on average, the residual power scales with volume, as
expected from the construction of the point source populations from
volume averaged luminosity functions. We also find that the
contaminating power increasing with redshift. This is related
to the adaptive smoothing scale described in
\sect~\ref{sec:radioanalysis:filter}, and can be intuitively
understood as it being harder to separate point sources from extended
emission at higher redshifts where the size of these two components
become more similar, given a constant resolution.

We find that the contaminating power is well fit by a power law
of the form
\begin{equation}
  \label{eq:biascorr}
  P_{\text{contamin}} = C ~ {r_{500}}^{3} ~ (1+z)^{\gamma},
\end{equation}
where $C$ is a normalization which has a slight dependence on the flux
extraction method used, and $\gamma=6.5\pm0.5$. We verify that the
same result is obtained regardless of whether the full simulation or
the point source-only simulation is used.

\begin{figure}
  \fontsize{14}{16}\selectfont
  \centering
  \includegraphics[width=\columnwidth]{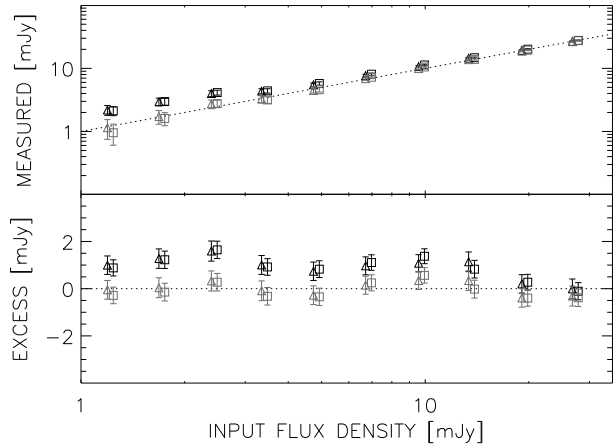}
  \caption{Measured ({\referee black}) and bias corrected ({\referee
      gray}) flux densities vs. input from simulations, using the
    radial fit ({\referee triangles}) and direct integration
    ({\referee squares}) methods of flux extraction. The measured
    values have been corrected for the finite aperture used
    (\sect~\ref{sec:sim:filter}), {\refereetwo while the bias corrected
      ones include an additional correction for} residual flux from
    point sources (\sect~\ref{sec:sim:ptsrc}).}
  \label{fig:fluxexcess}
\end{figure}

Correcting for the excess power, we further verify that the resulting
distribution of flux from a simulation is consistent with the input
flux, corrected for the finite aperture of the flux extraction
(Fig.~\ref{fig:fluxexcess}).

While the correction that we apply is a mean value, given a mass and
redshift, the actual power excess will vary. We account for the
scatter in the actual radio point source luminosity function in
clusters as an additional systematic uncertainty. We use the standard
deviation of excess signal from the simulations described above and
add this quantity in quadrature to the measurement uncertainty from
the filtered map.

\subsection{Radio lobes}
\label{sec:sim:lobes}

We now consider whether extended radio lobes can contribute to the
total derived extended radio luminosity. To this end, we extract
cutouts from the VLA FIRST survey \citep{1995ApJ...450..559B} at the
same frequency. The FIRST survey sky coverage is different from that
of NVSS, and we find counterparts for about one-third of the target
fields. For the latter, we visually inspect the FIRST maps, which have
a resolution of 5$^{\prime \prime}$, and find radio lobes around
bright sources within $r_{500}$ in more than half of the fields
inspected. We find that virtually all radio lobes found in the FIRST
maps are well within the regions that were flagged around bright NVSS
sources due to our filtering algorithm
(cf. \sect~\ref{sec:radioanalysis:filter}), and thus conclude that
unresolved radio lobes cannot contribute a significant fraction to our
derived radio signals.

\subsection{Null tests}
\label{sec:nulltest}

\begin{figure} 
  \centering
  \includegraphics[width=\columnwidth]{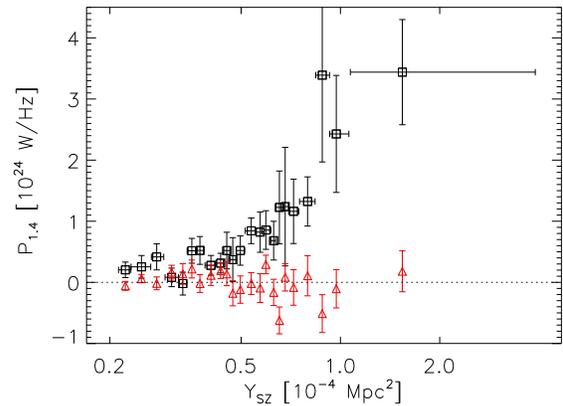}
  \caption{Null test for the PSZ(V) sample, with radio luminosity binned
    by Comptonization. The radio luminosities derived from the actual
    PSZ targets are shown as black squares with error bars, where the
    horizontal bars represent bin widths. Similarly binned radio
    luminosities derived from fields not \textit{a priori} associated with
    clusters of galaxies are shown as red triangles.}
  \label{fig:nulltest}
\end{figure}

In order to ascertain that there is no systematic contamination from
point sources or map filtering other than those discussed in the
above, we perform a series of null tests, associating random positions
in the sky with the PSZ and X-ray targets. After obtaining the
corresponding NVSS images and performing the complete analysis exactly
as for the real targets, we bin the resulting measured radio
luminosities (not bias-corrected {\referee in the sense of \eqtn~(\ref{eq:biascorr})}, since there is no \textit{a priori}
association with any massive cluster of galaxies) by $Y_{\text{SZ}}$
and $L_{\mathrm{X}}$ and compare to the on-target results. In general,
we find no excess flux in either of thus generated data set, as shown
for the PSZ(V) sample in Figure~\ref{fig:nulltest}. We also bin the
measured radio luminosities by the angular area on the sky
corresponding to $r_{500}$ and again find no excess flux.

Additionally, we test for a possible offset of the measured signal in
the vicinity of bright point source. Such bias can be caused by
inaccuracies in the removal of compact sources directly in the maps
(\sect~\ref{sec:radioanalysis:filter}) or by so-called clean bias
\citep{1997ApJ...475..479W,1998AJ....115.1693C}. Using randomly chosen
NVSS fields (as in the previous null test), we select 200 fields with
compact sources brighter than $500$~mJy, centering each field upon the
bright compact source, before applying our filtering
algorithm. Finally we stack the cleaned maps in two ways: with uniform
weighting (to test whether there is a flux bias independent of the
peak flux) and weighting by the peak flux (to test whether there is a
bias dependent on the magnitude of the compact source). Although we do
find a slight bias in the uniformly weighted case, the total
(integrated) bias level inside an aperture of one arc-minute is at a
level of less than $20~\mu$Jy/beam and is thus of no consequence for
our results. We find no evidence that the bias scales with the peak flux of
the compact source, thus there is no need to consider the peak
flux-weighted case separately.

\subsection{Individual flux comparison}
\label{sec:indiv:halo}

\begin{figure} 
  \centering
  \includegraphics[width=\columnwidth,trim=0 17 0 0,clip=true]{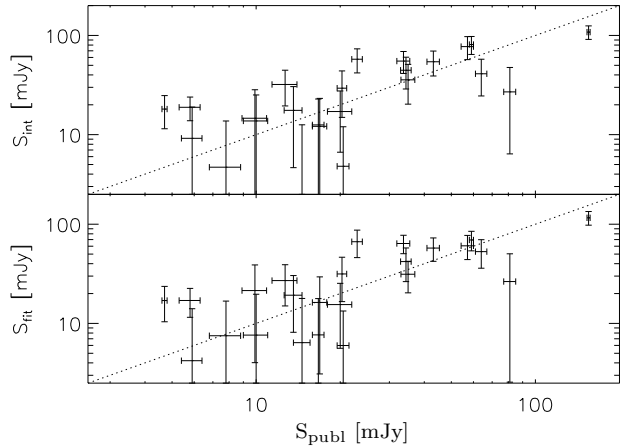}
  \hspace*{0.8cm}$\mathrm{S}_\mathrm{publ}$ [mJy]
  \vspace*{0.2cm}
  \caption{Comparison of integrated flux density measurements for a
    sub-sample of the X-ray clusters with the published measurements
    of \citet{2009A&A...507.1257G} and \citet{2012A&ARv..20...54F}
    {\referee denoted collectively here as $S_{\mathrm{publ}}$}. Flux density
    measurements using the direct integration method of this work have
    been corrected for the finite aperture used for integration
    (\sect~\ref{sec:sim:filter}).}
  \label{fig:gcomp}
\end{figure}

{\referee We conclude this section by comparing individual radio halo
  flux measurements with previously published results where the latter
  are available.}  The published data are collected from
\citet{2009A&A...507.1257G} and \citet{2012A&ARv..20...54F} (the
latter sample containing the former save for one target).
One goal of this comparison is to determine whether
the limited $uv$-coverage of the NVSS snapshots cause any systematic
flux attenuation of these diffuse sources.
We carry out the comparison {\referee for the union of all our samples}, for both
flavors of flux extraction after correcting the directly integrated
signals for the attenuation effect due to the finite integration
aperture, as described in \sect~\ref{sec:radioanalysis:directint}. The
result of the comparison is summarized in Fig.~\ref{fig:gcomp}.

We note that while the two flux extraction methods used in this study
yield consistent results, there are several cases in which we find
results very different from the published values. Considering the
different approaches to measuring the radio signals, this is
expected. While the results quoted by \citet{2009A&A...507.1257G} and
\citet{2012A&ARv..20...54F} rely on the visual identification of radio
halos and subsequent measurements of the flux within a thus defined
aperture, our measurements are based on the assumption of a universal
profile for the radio signal, and in addition we do not attempt to
visually distinguish radio halos from other potential sources of
extended emission (such as mini-halos or radio relics). These effects
lead to our method yielding fluxes significantly greater than the
previously published radio halo fluxes in many cases.

For other targets (a case in point is Abell 2163) the radio emission
is significantly more peaked than the mean profile derived in
\sect~\ref{sec:radioanalysis:profile}, which leads us to underestimate
the flux when using the fitting method. We also note that the NVSS
maps are often plagued by side lobe structures and artifacts,
especially where there is a bright, peaked radio emission is
present. This is the case for Abell 2219, where we recover less than
$50\%$ of the flux quoted by \citet{2012A&ARv..20...54F}. We also note
that our measurements are sensitive to the chosen centroids, which in
this work are simply those extracted from the respective cluster
catalogs. This can be expected to cause an underestimation in flux with
respect to published results, in particular for low-mass and
high-redshift targets, where the resulting linear scale is smaller.

{\referee A quantitative comparsion of the recovered flux in our two
  main samples is a potential way of testing for possible biases in
  the results presented in the next section. In practice, however,
  this is a difficult test to perform from the available public
  data. Firstly, published radio halo data are mostly ad-hoc
  compilations, so the resulting sample will have no specific
  selection function. In addition, there are no public radio halo data
  from SZ selected clusters. Secondly, for the objects shared in
  common between the union of our samples and
  \citet{2012A&ARv..20...54F}, most are also shared between the PSZ
  and the X-ray samples (23 out of 28 clusters), and therefore the
  resulting mean flux ratios between our analysis and the published
  data are thus almost identical for the two sub-samples. We conclude
  that although we are unlikely to have systematic differences in the
  flux recovery between the PSZ and X samples, a direct comparison
  with published radio halo data presently does not offer a conclusive
  test for such a bias.}


\section{Results}
\label{sec:results}

We now have all the tools in place to approach the main objective of
our work, namely to find the correlations between radio halo power and
the mass observables ($Y_{\mathrm{SZ}}$ and $L_{\mathrm{X}}$) in our
cluster samples and quantify the fraction of clusters that do not
follow the main correlation (i.e. consistent with having no radio halo
emission).  We carry out the regression analysis on the measured radio
luminosities found by the two methods described in
\sect~\ref{sec:radioanalysis:fluxextraction} {\referee and present the
  results in \sect~\ref{sec:results:model}. We compare our results to
  the literature in \sect~\ref{sec:res:compare}, and discuss the
  direct scaling of mass and radio halo power in \sect~\ref{sec:res:p-m}.}

\begin{table*} 
  \caption{Best-fit (maximum likelihood) parameters of the {\refereethree power law $P_{\text{1.4~GHz}} = A_{\mathrm{lim}} ( \frac{x}{x_{\mathrm{lim}}} ) ^B$, where $x$ is the mass observable ($Y_{\mathrm{SZ}}$, measured in $10^{-4}~\text{Mpc}^{2}$, 
      or $L_{\mathrm{X}}$, measured in $10^{44}~\mathrm{erg} ~ \mathrm{s}^{-1}$). $A_{\mathrm{lim}}$ and $x_{\mathrm{lim}}$ are the re-normalized regression parameters as defined in the text.}
    For each cluster sample, the analysis was carried out for the two different flux
    extraction methods described in the text, and fitted using the mixture
    model for the dropout population.}
\label{tab:fitres}      
\centering  
\begin{tabular}{l l l l l l l l}
& & & & & \\
\multicolumn{8}{c}{Variable mass limit} \\
\hline \hline
Sub-sample   & Mass  & Sub-sample & Flux extraction & Normalization    $A_{\mathrm{lim}}$                    & Power law  &  Intrinsic                   & dropout       \\ 
            & limit & size      & method          & [$10^{24} \mathrm{W} \, \mathrm{Hz}^{-1}$] & slope $B$  &  scatter $\sigma_{\mathrm{F}}$  & fraction $g$  \\
\hline
X(V) & $z$-dependent &   85 &         Radial fit & $  0.56^{+0.49}_{-0.24}$ & $  1.13^{+0.27}_{-0.17}$ & $  0.98^{+0.45}_{-0.20}$ & $  0.51^{+0.09}_{-0.13}$ \\ 
X(V) & $z$-dependent &   85 & Direct integration & $  0.37^{+0.43}_{-0.24}$ & $  1.48^{+0.53}_{-0.30}$ & $  0.54^{+0.40}_{-0.17}$ & $  0.70^{+0.07}_{-0.11}$ \\ 
PSZ(V)          & $z$-dependent &   89 &         Radial fit & $  0.52^{+0.26}_{-0.12}$ & $  1.63^{+0.24}_{-0.28}$ & $  0.82^{+0.39}_{-0.15}$ & $  0.25^{+0.11}_{-0.13}$ \\ 
PSZ(V)          & $z$-dependent &   89 & Direct integration & $  0.37^{+0.19}_{-0.10}$ & $  1.75^{+0.32}_{-0.22}$ & $  0.86^{+0.42}_{-0.17}$ & $  0.29^{+0.12}_{-0.12}$ \\ 
\hline
& & & & & \\
\multicolumn{8}{c}{Constant mass limit} \\
\hline\hline 
Sub-ample & Mass   & Sub-sample & Flux extraction & Normalization  $A_{\mathrm{lim}}$                   & Power law  &  Intrinsic                   & dropout       \\ 
            & limit  & size      & method          & [$10^{24} \mathrm{W} \, \mathrm{Hz}^{-1}$] & slope $B$  &  scatter $\sigma_{\mathrm{F}}$  & fraction $g$  \\
\hline
X(C) & $8 \times 10^{14} M_{\odot}$  &   77 &  Radial fit         & $  0.33^{+0.68}_{-0.28}$ & $  1.55^{+0.54}_{-0.48}$ & $  0.77^{+0.30}_{-0.29}$ & $  0.58^{+0.12}_{-0.11}$ \\ 
X(C) &  $8 \times 10^{14} M_{\odot}$ &   77 &  Direct integration & $  0.51^{+0.93}_{-0.45}$ & $  1.31^{+0.65}_{-0.52}$ & $  0.71^{+0.39}_{-0.22}$ & $  0.71^{+0.06}_{-0.11}$ \\ 
PSZ(C)         & $8 \times 10^{14} M_{\odot}$  &   79 &  Radial fit         & $  0.56^{+0.19}_{-0.12}$ & $  1.59^{+0.26}_{-0.23}$ & $  1.14^{+0.26}_{-0.16}$ & $  0.11^{+0.08}_{-0.11}$ \\ 
PSZ(C)         & $8 \times 10^{14} M_{\odot}$  &   79 &  Direct integration & $  0.50^{+0.18}_{-0.11}$ & $  1.63^{+0.28}_{-0.20}$ & $  1.06^{+0.25}_{-0.14}$ & $  0.18^{+0.10}_{-0.11}$ \\ 
\hline
\vspace{-2mm}
$~$ \\
\end{tabular}
\end{table*}

\subsection{Model fit}
\label{sec:results:model}

The results of our likelihood analysis are summarized in
\tabl~\ref{tab:fitres}.  {\refereethree In order to have physically
  meaningful and internally comparable values for the power law
  normalization, we re-scale Equation \ref{eq:powerlaw:general} as
\begin{equation}
  y = A_{\mathrm{lim}} \left( \frac{x}{x_{\mathrm{lim}}} \right) ^B.
\end{equation}
The newly defined power-law normalization, $A_{\mathrm{lim}}$,
corresponds to the value $x_{\mathrm{lim}}$ of the mass observable,
obtained using the definitions in Section \ref{sec:sel}, for a fixed
mass of $M_{500} = 6\times 10^{14} M_{\odot}$. Thus, for the X-ray
subsamples, $x_{\mathrm{lim}} = 3.6\times 10^{44} ~\mathrm{erg}
~\mathrm{s}^{-1}$, and for the SZ subsamples $x_{\mathrm{lim}} =
0.43\times 10^{-4} ~\mathrm{Mpc}^2$}.

The relatively larger uncertainties in the best-fit
parameters for the X-ray sub-samples, in spite of sample sizes similar
to PSZ, reflect a higher dropout fraction, and thus fewer data points
serving to constrain the power law.

\begin{figure*} 
  \includegraphics[width=0.94\columnwidth]{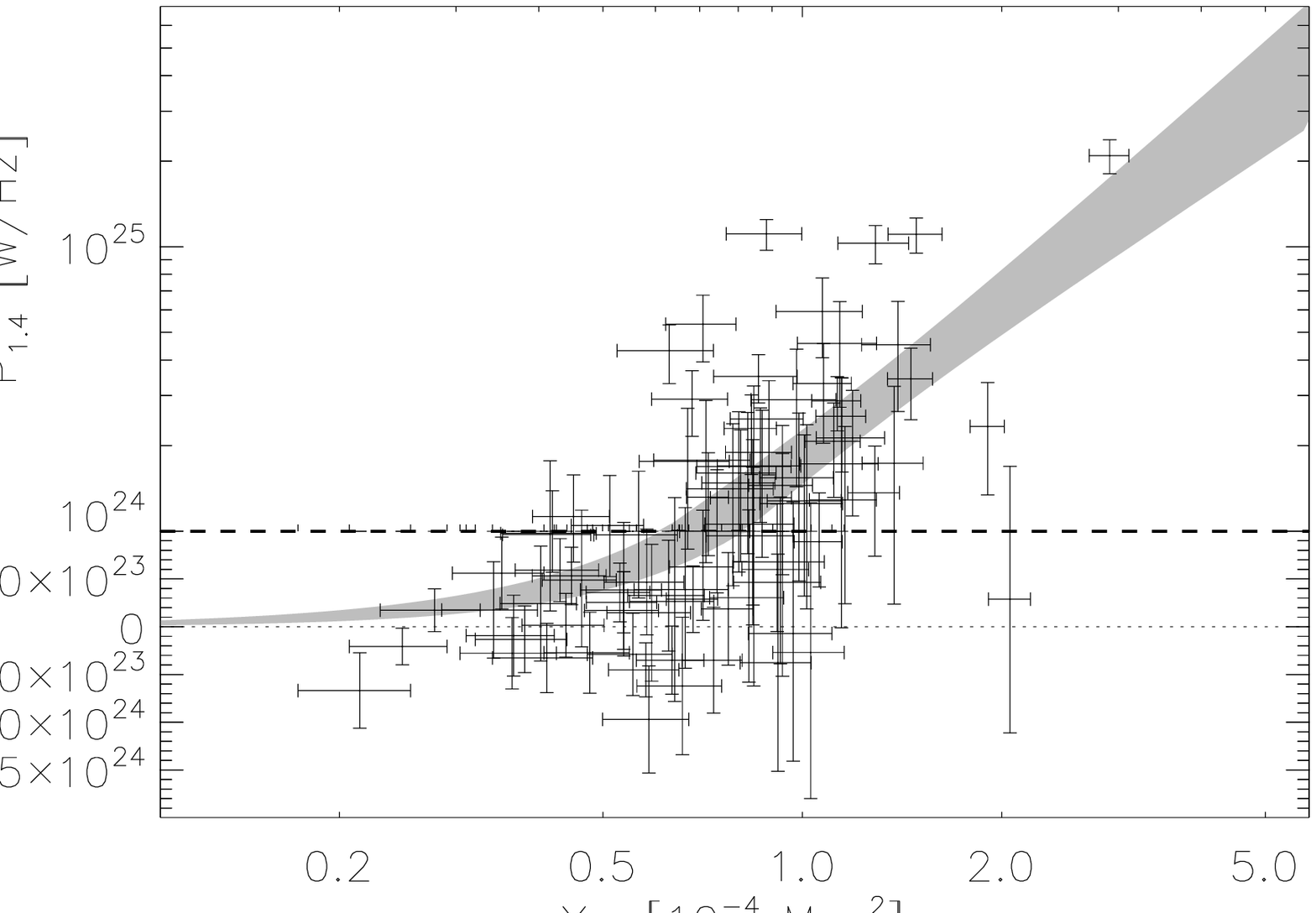}%
  \hspace*{6mm}
  \includegraphics[width=0.94\columnwidth]{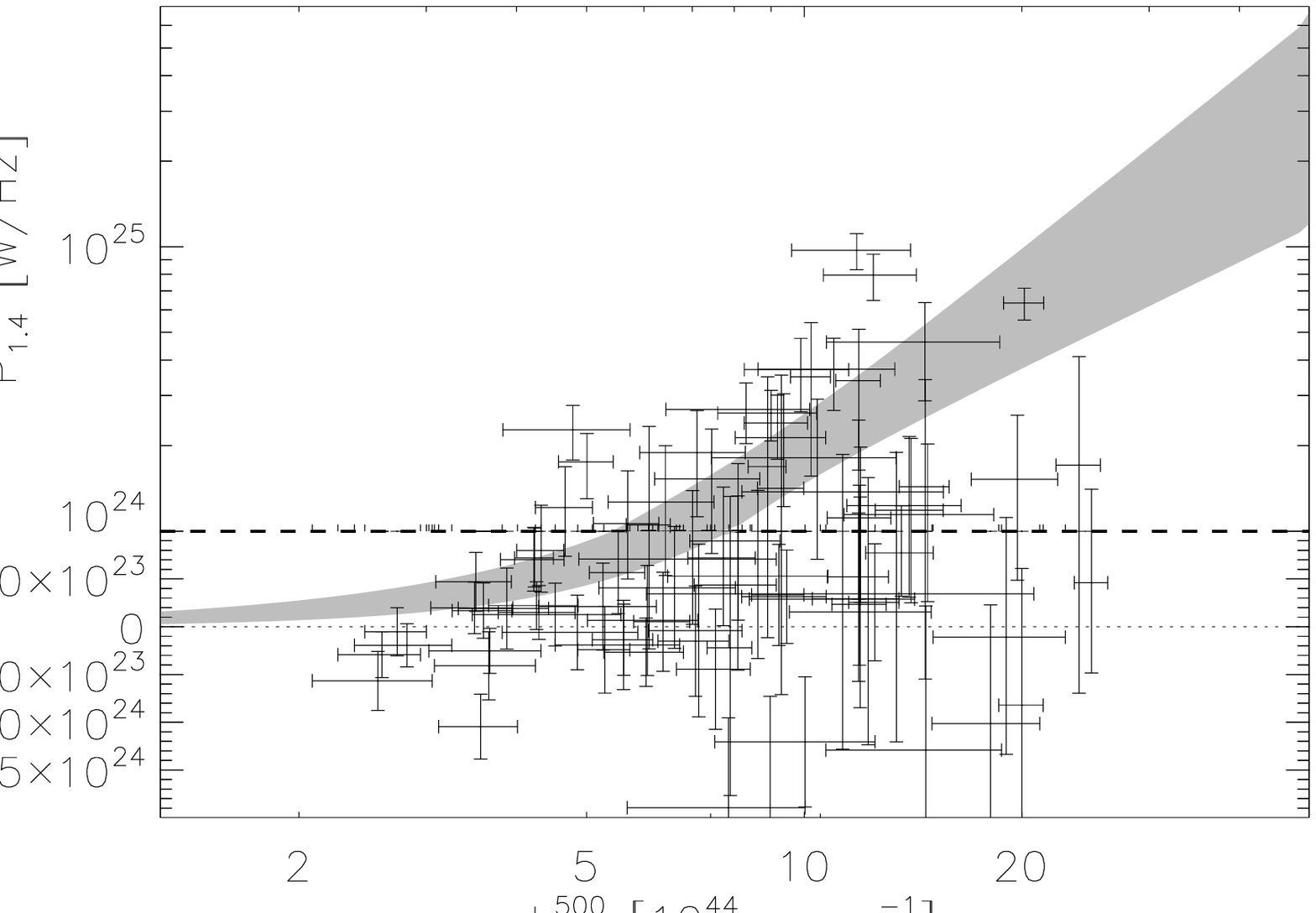}%
  
  \includegraphics[width=0.94\columnwidth]{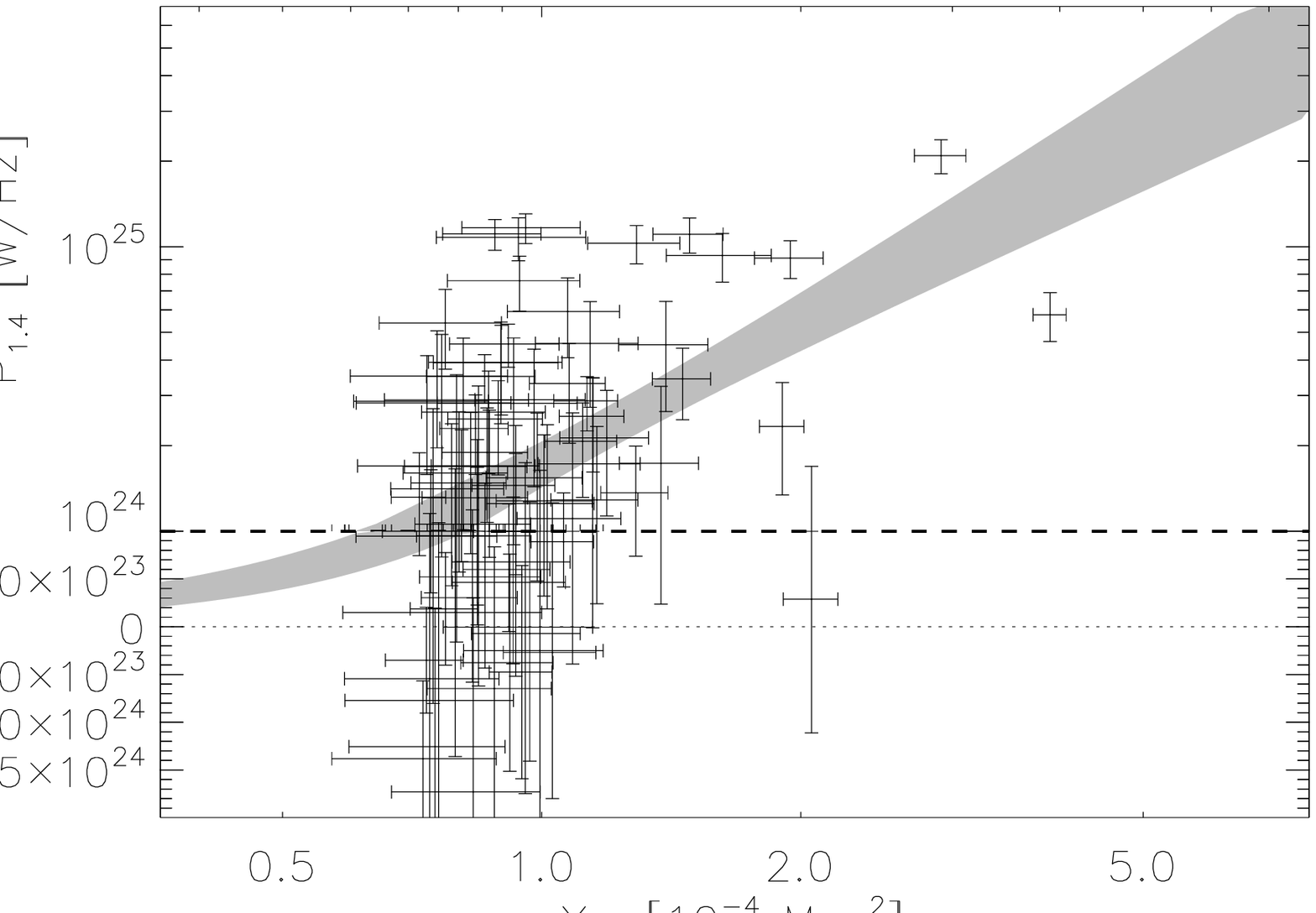}
  \hspace*{6mm}
  \includegraphics[width=0.94\columnwidth]{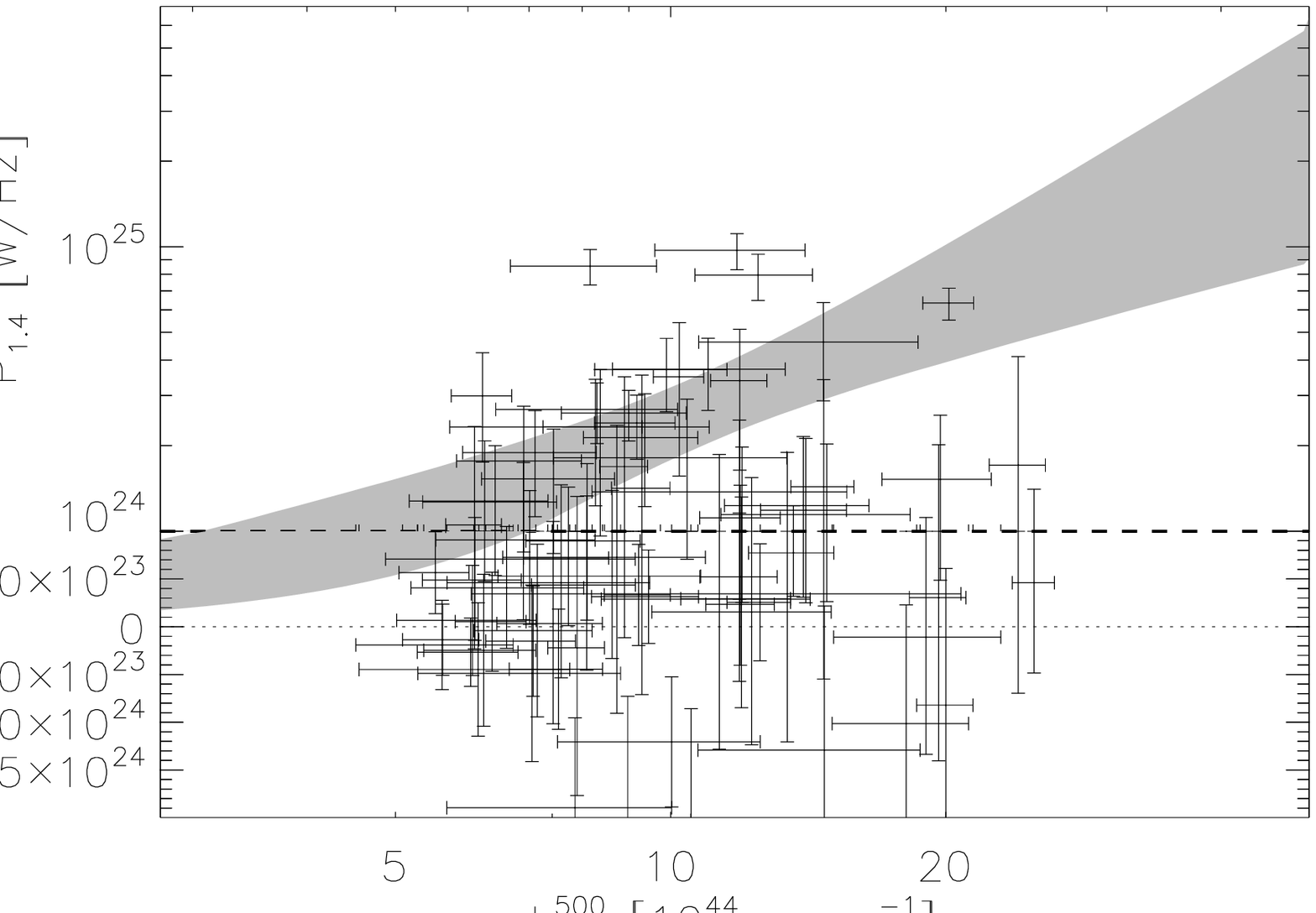}

  \caption{ {\tempbf 1.4 GHz radio power plotted against
      $Y_{\text{SZ}}$ for the PSZ (\textit{left column}) and X-ray
      (\textit{right column}) sub-samples, using the direct
      integration method for flux extraction. The \textit{upper
        panels} show the case for a variable mass limit (PSZ(V) and
      X(V), respectively), and the \textit{lower panels} are for a
      constant mass-cut (PSZ(C) and X(C), respectively).}  The gray
    shaded region in each panel represents the 68\% confidence region
    of the {\refereetwo best-fit power-law, using the mixture model to
      account for the off-population as described in the text}. Note
    that the ordinate axes are broken into linear and logarithmic
    parts.}
  \label{fig:esz}
\end{figure*}

\begin{figure} 
  \centering
  \includegraphics[width=\columnwidth]{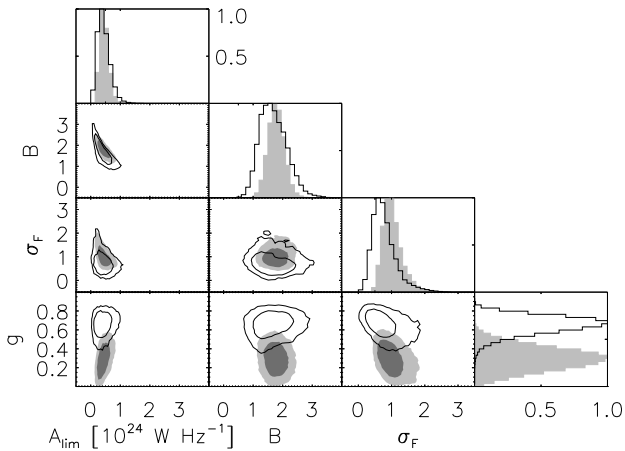}
  \caption{Marginalized posterior probability distributions for the
    maximum-likelihood analysis for the sub-samples with redshift
    dependent mass limits. Filled (gray) and black contours represent
    the results for the PSZ(V) and X(V) sub-samples {\refereetwo
      (direct integration method)}, respectively. The contours
    encompass 68\% and 95\% of all sampled points in the posterior,
    marginalized over the other parameters. Also shown are histograms
    representing the (normalized) marginalized distributions of single
    parameters of the fit.}
  \label{fig:vbanana}
\end{figure}

\begin{figure} 
  \centering
  \includegraphics[width=\columnwidth]{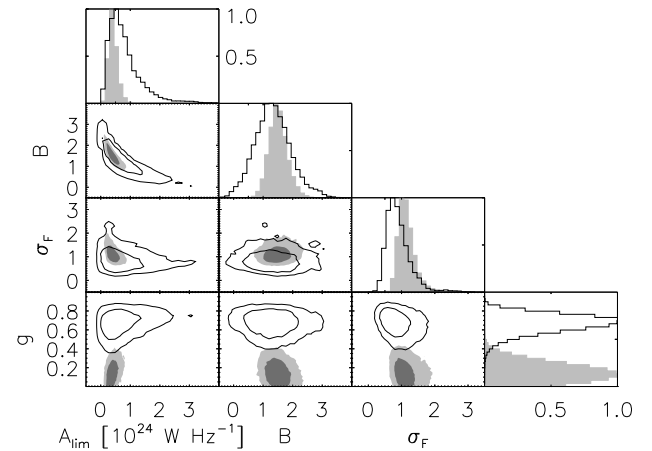}
  \caption{As figure \ref{fig:vbanana}, but for the PSZ(C) (gray,
    filled contours) and X(C) (solid black contours)
    sub-samples, selected by a constant mass limit.}
  \label{fig:cbanana}
\end{figure}

{\tempbf In Figure \ref{fig:esz} we show} the radio power versus the
mass observable for the different sub-samples, accompanied by the
allowed range (at 68\% confidence) of power law models. Although the
effect of contamination from radio point sources is small, we model
the power law and dropout-fraction taking this effect into account as
described in \sect~\ref{sec:sim:ptsrc}. It is clear that most of our
radio halo measurements are formally non-detections ({\refereetwo the fraction of measurements above
    a $3\sigma$ threshold being in the approximate range of
    20\%-35\%}). We refer to \sect~\ref{sec:indiv:halo} for a
  quantitative comparison with published results.
We visualize the posterior likelihood distributions by marginalizing
over the four regression parameters (the slope $B$ and normalization
$A_{\mathrm{lim}}$, the intrinsic scatter in the radio power
$\sigma_F$, and the dropout fraction $g$), as displayed in Figures
\ref{fig:vbanana} and \ref{fig:cbanana}. 

The model parameters for the power law are essentially consistent
between the different samples. This is expected, since the
core-excised soft band X-ray luminosity scales almost linearly with
the integrated SZ signal \citep[e.g. $Y_{\mathrm{SZ}} \propto
L_{\mathrm{X}}^{1.14\pm 0.08}$,][]{2010A&A...517A..92A}, and we do not
expect to see the small difference in the power-law coefficients given
the large statistical errors (see \sect~\ref{sec:res:p-m} for a
direct comparison in terms of mass scaling).  The dropout fractions,
however, are generally inconsistent between the SZ and X-ray selected
samples. This is especially the case comparing the constant-mass
selected sub-samples.

It is conceivable that differences in the redshift distribution of
clusters can cause the observed differences in the dropout fraction.
Although the redshift distributions of the PSZ and X-ray sub-samples
are very similar below $z<0.3$, as indicated in Fig.~\ref{fig:zhists},
the higher number of $z > 0.3$ objects in the PSZ selection is a cause
for concern. It remains a possibility that in the high redshift
universe the merging fraction is higher and hence the PSZ sub-samples
show an increased occurrence of radio halos. To test this, we
re-perform the analysis limiting the samples to redshifts below $z =
0.3$. Although the parameter uncertainties increase due to the lower
samples sizes, the results are in general agreement with the above. We
find $g=0.66^{+0.08}_{-0.22}$ and $g=0.25^{+0.11}_{-0.21}$ for the
X(V) and PSZ(V) samples, respectively, and $g=0.53^{+0.13}_{-0.20}$
and $g=0.00^{+0.11}_{-0.00}$ for the corresponding (C) samples. We
also find consistent slopes and normalizations in each case.
{\refereetwo Finally, we note that the intrinsic scatter is found to
  be very large (close to 100\%) in all samples, consistent
  with the estimates of B12.}

\subsection{Comparison with published results}
\label{sec:res:compare}

\begin{figure} 
  \centering
  \includegraphics[width=\columnwidth]{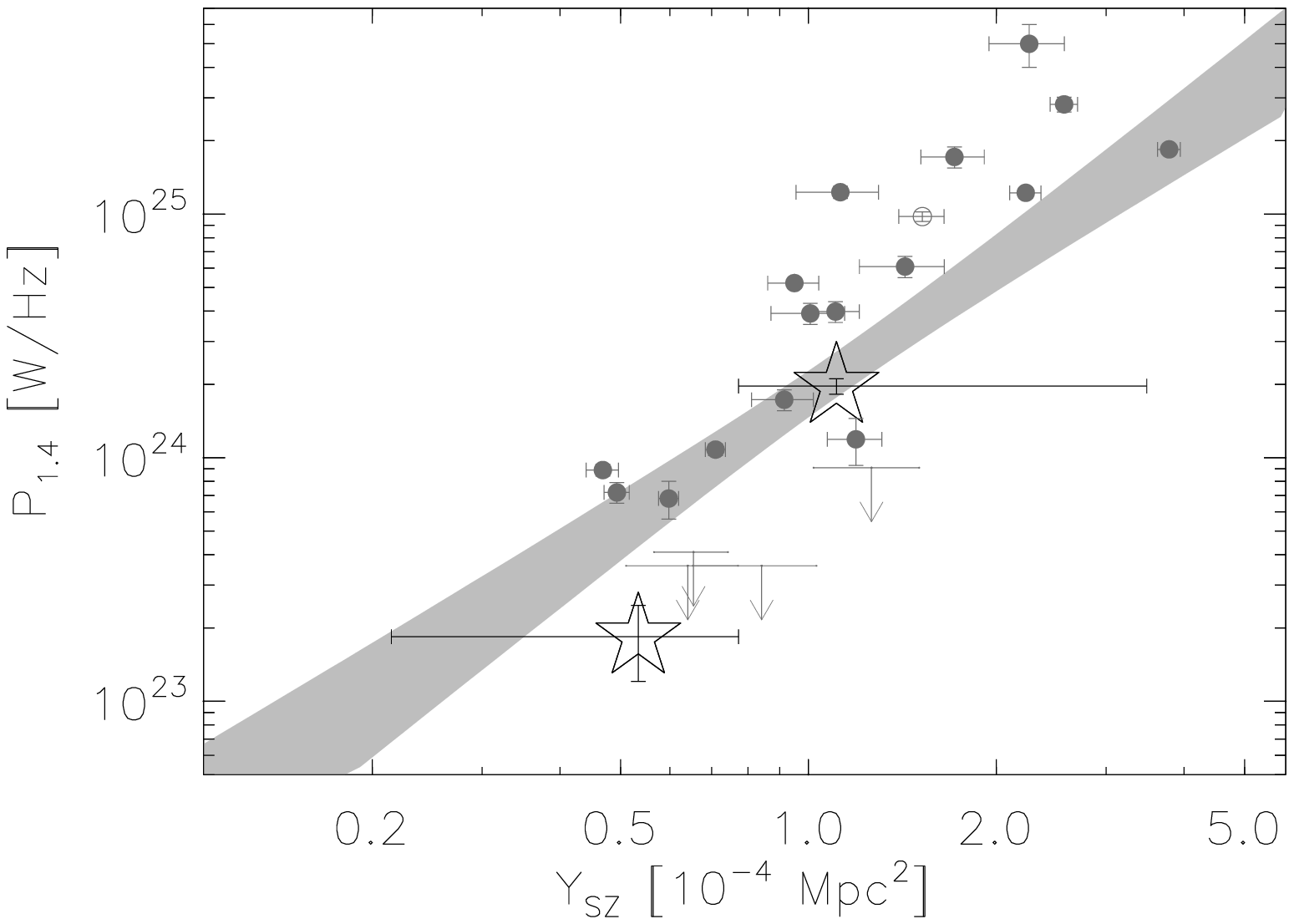}
  \caption{Weighted mean values (open stars) of $Y_{\text{SZ}}$ and
    $P_{1.4\text{GHz}}$ after dividing the PSZ(V) sample (direct
    integration method) into two bins equal by number of
    targets. The 68\% confidence region of the power law fit of this
    work is indicated in gray. Also indicated are previously known
    radio halos (filled circles), mini-halos (open circles) and upper
    limits (arrows), selected from \citet{2009A&A...507..661B} based
    on cross-selection with PSZ.}
  \label{fig:szcompar}
\end{figure}

We now compute weighted averages of the derived radio halo
luminosities and compare to published correlations with X-ray and SZ
mass observables. We divide each sample into two equally large (by
number of targets) mass bins, and average the radio luminosities
weighted by the inverse square of their uncertainties. In
Fig.~\ref{fig:szcompar} we compare our result from the PSZ sample with
the radio/SZ correlation of B12, based on the compilations of
\cite{2009A&A...507..661B} and \cite{2009A&A...507.1257G}.  As
expected, the mean values are generally consistent with the
correlation of the "radio-on" population of B12, since the dropout
fraction in the PSZ sample is small (or consistent with zero at the
high mass end). In the low mass bin the fraction of dropouts is
higher, hence the mean value lies below the 68\% confidence region of
the radio-halo only regression (indicated by the gray band in
Fig.~\ref{fig:szcompar}). This result again illustrates that there is
no net bias in the radio halo flux measurements from our method
compared to previous studies.

\begin{figure} 
  \centering
  \includegraphics[width=\columnwidth]{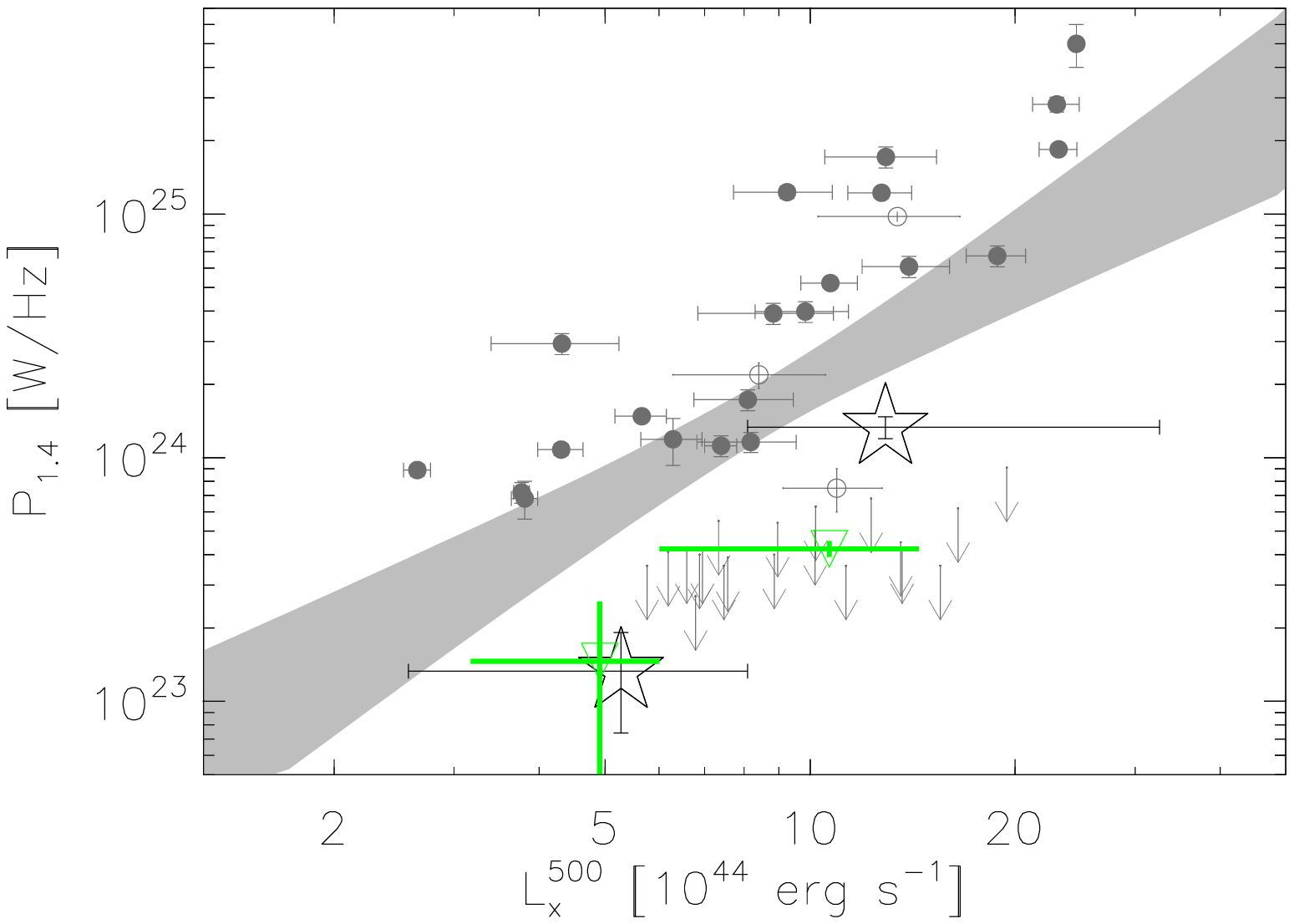}
  \caption{As Figure~\ref{fig:szcompar}, but for the X(V) sample
    {\referee (direct integration method)}. All radio halos (filled
    circles), mini-halos (open circles) and upper limits (arrows) of
    \citet{2009A&A...507..661B} are indicated. Also indicated are the
    measurements of \citet{2011ApJ...740L..28B} (green triangles) in
    two bins of X-ray luminosity.}
  \label{fig:xraycompar}
\end{figure}

For the correlation with $L_{\mathrm{X}}$ we again use the
\cite{2009A&A...507..661B} sample that includes radio halos,
mini-halos and upper limits on non-detections. We compare these
results to the mean signals in our two $L_{\mathrm{X}}$ bins
(Fig.~\ref{fig:xraycompar}), and find that they are below the 68\%
confidence interval of the radio halo correlation found in
\sect~\ref{sec:results}. This is expected since the mean values are an
average of the ``on" and ``off" populations (the latter being
approximately $\sim 70$\% in the X-ray selection). In
Fig.~\ref{fig:xraycompar} we also show the results of
\cite{2011ApJ...740L..28B}, from stacking of SUMSS radio maps in a
sample of X-ray clusters. We find that the signal in the
low-$L_{\mathrm{X}}$ bin from Brown et al.  is consistent with our
measurement, but in the higher $L_{\mathrm{X}}$ bin their upper limit
is significantly lower than the mean signal derived from our analysis.

Since the mean values from the stacked radio images of X-ray clusters
of \cite{2011ApJ...740L..28B} have been claimed as a detection of the
radio halo ``off-state", the factor $\sim$3 discrepancy with our mean
value in the high $L_{\mathrm{X}}$ bin requires some clarification. We
suggest this difference is likely caused by a combination of several
factors, both in the sample selection and map filtering
procedures. The Brown et al. sample is not X-ray flux limited and
hence their high-$L_{\mathrm{X}}$ sample may contain several lower
mass objects with peaked X-ray emission (see
\sect~\ref{sec:analysis}). Their sample is also cut to approximately
two-thirds of its original number by a criterion that relates the peak
flux in each field with cluster mass, and hence can systematically
remove some of the brightest radio halos as well as massive, cool-core
clusters with a central AGN. On the map filtering side, the use of the
multi-scale spatial filter of \cite{2002PASP..114..427R} can cause an
under-estimation of radio halo flux if emission is highly clumped
\citep[see the discussion in][]{2009ApJ...697.1341R}. We tested the
performance of the \cite{2002PASP..114..427R} filter on our simulated
maps and found it to be more strongly affected by residual point
source flux and substructures.  Finally, the use of a fixed filtering
scale of 600 kpc in Brown et al. for all clusters will create a bias
against smaller radio halos in low-mass objects (conversely, removed
the contribution of mini-halos that can potentially contaminate our
results). Based on these considerations we argue that the mean flux of
radio halos in X-ray luminous clusters might have been under-estimated
by Brown et al.

\subsection{$P_{\mathrm{1.4~GHz}}-M_{500}$ scaling}
\label{sec:res:p-m}

\begin{figure} 
  \centering
  \hspace*{-0.7cm}\includegraphics[width=\columnwidth]{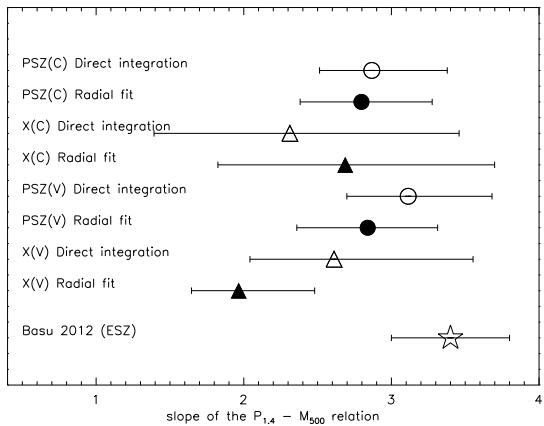}
  \caption{Slope of the $P_{\mathrm{1.4~GHz}}-M_{500}$ scaling:
    $P_{\mathrm{1.4~GHz}} \propto M_{500}^\gamma$, from the various
    flavors of sample selection and flux extraction methods. The B12
    result is included for comparison.}
  \label{fig:pmrel}
\end{figure}

Our measured scaling of the radio halo luminosities with mass
observables lead to consistent results between the SZ and X-ray
selected samples.  We estimate the mass scaling of radio halos by
measuring the slope of the $P_{\mathrm{1.4~GHz}}-M_{500}$ correlation
directly from the posterior distribution of
$Y_{\mathrm{SZ}}-P_{\mathrm{1.4~GHz}}$ and $L_{\mathrm{X}} -
P_{\mathrm{1.4~GHz}}$ slopes presented in the previous section, in
conjunction with the mass-observable scaling relations discussed in
\sect~\ref{sec:sel}. The results are presented in
\tabl~\ref{tab:pmscaling} and graphically in Figure~\ref{fig:pmrel}.

\begin{table} 
\caption{Slope of the $P_{\mathrm{1.4~GHz}}-M_{500}$ scaling: $P_{\mathrm{1.4~GHz}} \propto M_{500}^\gamma$ }       
\label{tab:pmscaling}      
\centering                 
\begin{tabular}{l l l}
\hline\hline               
Sub-sample     & Flux extraction & Power law  \\ 
sample        & method          & slope $\gamma$      \\
\hline
X(V) & Radial fit         & $1.96^{+0.50}_{-0.32}$ \\
X(V) & Direct integration & $2.57^{+0.97}_{-0.52}$ \\
PSZ(V)          & Radial fit         & $2.84^{+0.46}_{-0.50}$ \\
PSZ(V)         & Direct integration & $3.10^{+0.57}_{-0.40}$ \\
X(C) & Radial fit         & $2.68^{+0.98}_{-0.82}$ \\
X(C) & Direct integration & $2.78^{+1.13}_{-0.88}$ \\
PSZ(C)         & Radial fit         & $2.78^{+0.49}_{-0.40}$ \\
PSZ(C)         & Direct integration & $2.88^{+0.50}_{-0.36}$ \\
\hline
\end{tabular}
\end{table}

The current results are mostly consistent at one-sigma level with the
one found by B12 using an ad-hoc selection of known radio halos
present in the ESZ catalog ($P_{1.4} \propto M_{\mathrm{vir}}^{3.4\pm
  0.4}$).  There is, however, an indicative trend towards a
{\referee flatter} $P_{\mathrm{1.4~GHz}}-M_{500}$
correlation, as each of the sub-sample and flux estimation method
shows preference towards a smaller value than B12. The presence of
many high-mass clusters in the ad-hoc B12 selection and possible
absence of non-detections in the literature may be responsible for
this bias. We also note that the secondary or hadronic models predict
a shallower mass correlation of radio halo power than the primary or
re-acceleration ones
\citep[e.g.][]{2009JCAP...09..024K,2013ApJ...777..141C},
but we do not dwell further on this difference given its low
statistical significance. We also do not attempt to scale the
$Y_{\mathrm{SZ}}$ values from the PSZ catalog to inside the radio halo
radius (as was done in B12 to obtain a roughly linear correlation
between the SZ and radio signals) since neither of our flux extraction
methods allow a direct measurement of the radio halo radius from the
NVSS data.
 
The scaling of the radio halo power with the total cluster mass is a
more useful indicator for predicting radio halo counts from
cosmological-scale simulations. Such simulations are still in their
early stages, and the large number of free parameters need to be
matched against observations. \cite{2012ApJ...759...92S}, for example,
showed that the slope of the $P_{1.4}-M_{\mathrm{vir}}$ correlation
can be useful to distinguish different models of radio halo
origins. Even though the uncertainties in the scaling relation
presented in our work will not help in breaking most of the parameter
degeneracies, some models, e.g. those with high magnetic field
strengths, can be ruled out at high significance.

\section{Discussion}
\label{sec:discussion}

The main result of our analysis is a statistically significant
difference in the number of ``radio quiet" clusters in the SZ and
X-ray selected samples.  Having established that our radio flux
measurements are not subject to any map filtering bias or other
systematic artifacts, we turn our attention to a qualitative
understanding of this difference. We begin by considering biases
arising from the presence or absence of cool-core clusters in
different types of samples. Finding that such bias is unlikely to be
the sole contributor to the observed selection difference, we put
forward the different time evolution of the SZ and X-ray signals
during cluster mergers as a more likely cause. We offer some
predictions based on the latter hypothesis, and conclude this section
with a crude estimate of the expected number of radio halos in the
sky.

\subsection{Bias due to cool cores}
\label{sec:coolcore}

In our earlier work (B12) using the Planck ESZ catalog, we proposed
the over-abundance of cool-core clusters in X-ray selected samples as
a possible reason for strong bi-modality.  Cool core clusters are
predominantly relaxed systems which generally do not harbor giant
radio halos, although they often exhibit radio mini-halos at their
centers. The bias towards cool core clusters in X-ray selected samples
is well known. \cite{2011A&A...526A..79E} showed that up to $\sim
30\%$ of the strong cool-core objects should be removed from a given
X-ray flux-limited sample. Inclusion of several lower mass objects
near the mass limit of X-ray selected samples will thus create a bias
towards radio quiet systems.

An objection to the above argument is that a strong cool-core bias
will be more prominent at the low-$L_{\mathrm{X}}$ end, and thus the
massive clusters considered in our study should only be moderately
affected. {\referee The significant discrepancy in the radio halo
  hosting population between the X-ray and SZ selected clusters is
  difficult} to explain solely from the contamination of a few less
massive cool-core clusters near the X-ray mass selection threshold.

That being said, the appearance of cool cores will boost the X-ray
luminosity in a cluster disproportionate to its mass, thereby
``enhancing" the bi-modal division which is already prominent in the
X-ray selection. The recent results of \cite{2013ApJ...777..141C}
support this scenario: after excising the core emission in an X-ray
selected cluster sample these authors still find a bi-modal division
in the radio/X-ray correlation, but it is less prominent compared to
the core-inclusive values.  This result is fully consistent with our
finding of roughly 65\% radio dropouts in the X-ray selection. Note,
however, that the use of a lower scatter mass proxy (e.g. $Y_X$) in a
thus selected sample will lessen the visual perception of bi-modality.

Conversely, it can be argued that the PSZ cluster catalog is biased
\textit{against} cool-core systems due to their radio AGN
contamination. Cool core clusters tend to be associated with a radio
AGN at their center
\citep[e.g.][]{2009ApJ...704.1586S,2009A&A...501..835M}, which in
principle can offset or completely cancel the SZ decrement. However,
at the high S/N level that our PSZ clusters are selected such a
contamination is expected to be negligible. The PSZ catalog does
contain several high-mass, cool core systems, and the Planck team
identifies only 11 clusters with a central AGN that are present in the
MCXC catalog but not in the PSZ selection
\citep{2013arXiv1303.5089P}. Of these only 5 objects satisfy our
selection criteria, a small fraction compared to our sample sizes. It
is more likely that the expected SZ signals from these objects are
erroneously over-predicted based on their high $L_{\mathrm{X}}$
values, rather than the SZ signal being contaminated by a radio
AGN. In general, it has been shown from simulations
\citep{2010ApJ...709..920S} and direct observations
\citep{2013ApJ...764..152S} that radio AGN contamination in SZ surveys
is well below 5\% for the high-significance detections.  Thus we can
rule out any significant bias against cool core objects in the high
signal-to-noise PSZ clusters used in our study.

We conclude that even though the X-ray selected sub-samples can contain
several lower mass, X-ray luminous cool core clusters, these objects
alone are unlikely to create the significant difference in the radio
halo occupancy that we find between SZ and X-ray selection. If the
mass selection threshold is decreased further, the cool core bias will
create an increasing dropout fraction in the X-ray selection
\citep[which is indeed observed, see e.g.][]{2013A&A...557A..99K}, but
at the high-mass end its impact will be less severe.  To understand
our results, we therefore take a closer look at how the SZ and X-ray
mass observables change with time during mergers when the radio halos
are supposedly at their brightest.

\subsection{X-ray and SZ signal during mergers}
\label{sec:analysis}

\begin{figure}
  \centering
  \hspace*{0.1cm}\includegraphics[width=\columnwidth]{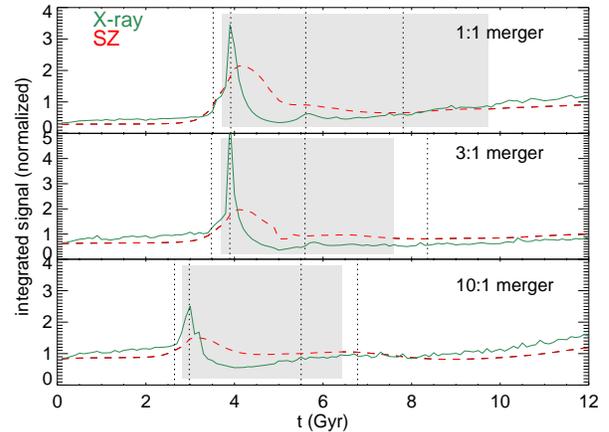}
  \caption{Relative changes in SZ and X-ray observables during cluster
    mergers \citep[simulation data from][]{2007MNRAS.380..437P}. The
    red-dashed and green-solid lines denote $Y_{\mathrm{SZ}}$ and
    $L_{\mathrm{X}}$ inside $r_{500}$, respectively. The signals are
    normalized to their final equilibrium values computed from the
    observed scaling and final mass.  The vertical dotted lines, from
    left to right, denote the times of virial crossing, first core
    crossing (first pericenter), second core crossing (second
    pericenter), and the time when the remnants will appear as a
    relaxed object in X-rays. The gray shaded regions are meant as
    tentative illustrations of radio halo lifetimes during mergers.}
  \label{fig:mergers}
\end{figure}

Simulations indicate that the merger related boost in the SZ signal is
much smaller than that of the X-ray signal. We use the simulation data
of \cite{2007MNRAS.380..437P} as an illustration of this point,
showing in Fig.\ref{fig:mergers} the time variation of the integrated
X-ray bolometric luminosity and the SZ signal inside $r_{500}$ during
and after mergers. The examples are for head-on mergers, with mass
ratios 1:1. 3:1 and 10:1, respectively. The SZ and X-ray signals are
normalized with respect to their final values, scaled from the initial
ones through observed correlations \citep[see][for
details]{2007MNRAS.380..437P}.  These controlled merger simulations
highlight the less severe fluctuations of the SZ signal during and
after the merger process, and its tendency to remain close to the
predicted scaling values in the subsequent relaxed phase (red lines in
Fig.~\ref{fig:mergers}).  More importantly for our purpose, the
integrated SZ signal, in contrast to the X-ray luminosity, does not
show a drop in intensity right after the cluster core passage. The
drop in central density in a disruptive merger is compensated by a
rise in the gas temperature, keeping the pressure close to its
equilibrium value.

Under the assumption that radio halos are fueled by post-merger
turbulence and energy dissipation, the drop ($\sim$50\% or more) in
the X-ray luminosity is likely to create a bias against radio halo
clusters in X-ray selected samples. We note that the $L_{\mathrm{X}}$
de-boost phase can last several Gyr, possibly covering the entire
radio halo lifetimes. On the other hand, there is an indicative bias
{\it towards} finding radio halos in SZ selection, as the delayed
boost in the SZ signal during mergers can preferentially aid the
detection of radio halo clusters. {\refereetwo However, the latter
  bias should be small if the duration of the radio halo is longer
  than the typical timescale of the SZ extended boost, which is on the
  order of 1 Gyr}.

The opposing trend between cluster X-ray luminosity and the radio halo
activity has also been demonstrated recently through high-resolution
MHD simulations \citep{2013MNRAS.429.3564D}, and we identify this as a
principal cause for the observed selection difference between
high-mass SZ and X-ray clusters. For illustration, we indicate
tentative radio halo lifetimes in Fig.~\ref{fig:mergers} as gray
shaded regions, where the lifespans are chosen to be proportional to
the total mass (we emphasize that this illustrative example is not
based on any simulation results).  An immediate consequence of the
above argument is the prediction of radio halos in X-ray
under-luminous late-merger clusters. Indeed such clusters have been
reported \citep[see][]{2011A&A...530L...5G}, but the numbers are
small, as expected, since the selection was done in X-rays. As blind
detections of radio halos from future radio surveys will be
followed-up through X-ray and SZ observations, the radio X-ray
correlation is thus expected to broaden significantly, while the radio
SZ correlation will remain less affected.

The rapid rise in X-ray luminosity, lasting for only $\sim 0.5$~Gyr,
can also explain those rare merging systems that do not host radio
halos, for example Abell 2146 \citep{2011MNRAS.417L...1R}. If the
onset of giant radio halos is not instantaneous and the radio
brightness reaches its maximum after the cluster core crossing (second
vertical dotted line from left in Fig.~\ref{fig:mergers}), then for a
short fraction of their life clusters can be significantly X-ray
bright but radio under-luminous. Again, the same offset will cause a
less severe scatter in the radio-SZ correlation since the boost in the
SZ signal during mergers is less prominent, and occurs at a slightly
delayed time-scale that potentially corresponds better to the peak in
the radio halo flux.

\subsection{Implications for future radio surveys}
\label{sec:implications}

Radio halos are challenging to detect from blind radio observations:
they are diffuse, low-brightness objects with considerable
sub-structure, often confused with radio relics and even extended
radio galaxies.  At low redshifts their Mpc scale emission cannot be
imaged with most radio interferometers, and at high redshifts their
surface brightness drops rapidly due to cosmological dimming and the
K-correction. Nevertheless, several upcoming low-frequency radio
surveys, such as LOFAR\footnote{http://www/lofar.org} and the SKA
pathfinder experiments which will become operational in the next
decade, can address these observational challenges. At 1.4 GHz, the
ASKAP/EMU survey \citep{2011PASA...28..215N} will map the radio sky
with roughly 10$^{\prime\prime}$ resolution and 40 times better
sensitivity than the NVSS, and a similar capability is also expected
from the WODAN project \citep{2011JApA...32..557R}.

Consequently there is a revived interest in observing radio halos (and
relics) with these instruments, but the theoretical predictions are
uncertain. Based on the observed low number count of radio halos in
X-ray selected clusters, \cite{2012A&A...548A.100C} predicted
{\refereetwo up to} $100-200$ objects in the whole sky with 1.4 GHz
fluxes $\gtrsim$ 1 mJy. There is much uncertainty in these numbers as
there are several model parameters that can vary by a wide margin
\citep{2010A&A...509A..68C}. There is a similar situation in the
cosmological simulations for radio halo formation
\citep{2012ApJ...759...92S}, which are at an early stage but tend to
predict more radio halos than re-acceleration based models. In light
of this, it is interesting to make a rough estimate of the radio halo
count in the entire sky based on our dropout fraction in SZ selected
samples.

At the constant mass cut used for the PSZ(C) and X(C) samples,
$M_{500} > 8\times 10^{14}$ M$_{\odot}$, there are over 200 clusters
in the entire sky \citep[computed using the mass function
of][]{2008ApJ...688..709T}, with the majority below redshift
$z=0.6$. This number increases to over 1800 for a mass cut of $M_{500}
> 5\times 10^{14}$ M$_{\odot}$, corresponding to the low-$z$
completeness limit of the Planck cosmological sample. In the redshift
range $z<0.5$ there are roughly 1000 clusters in the entire sky, and
Planck is expected to have most of these massive objects in its final
data release. If we use the dropout fraction from the PSZ(V) sample,
$g = 0.25 \pm 0.12$ using the radial fit method, we can expect roughly
$750 \pm 120$ radio halos in the sky in this redshift range. This is
 {\refereetwo roughly a factor of 5 more} 
than the current predictions at 1.4
GHz for the ASKAP/EMU and WODAN surveys \citep{2012A&A...548A.100C}.

The crucial information needed to make a more realistic prediction of
radio halo counts in the sky is how the dropout fraction varies with
cluster mass. We have been unable to probe the lower mass domain as
our selection is based on the Planck PSZ catalog of the most massive
objects in the universe. The noisy NVSS data also prohibits such an
analysis as the radio power decreases rapidly with cluster mass. These
intermediate- to low-mass objects will form the bulk of radio halo
detections in the future radio surveys at 1.4 GHz and at low
frequencies, so future work must address this mass-dependence issue
observationally by dedicated follow-up observation of several tens of
clusters. Finally, if a large fraction of clusters are found to be
host to radio halos at all masses, they can effectively be used to
probe the increasing merger rate of cluster-size dark matter halos
through the cosmic time, an important test for the concordance model
of cosmology.


\section{Summary and conclusions}
\label{sec:conclusions}

We have measured the rate of occurrence of radio halos in galaxy
clusters and their correlation with cluster mass observables. We
constructed two main cluster samples: an X-ray selected sample based
on the REFLEX, BCS/eBCS, NORAS and MACS catalogs, and an SZ selected
sample based on the Planck PSZ catalog. The cuts in the X-ray
luminosity and the integrated Comptonization parameters were chosen to
ensure near identical mass limits. The samples were cross-correlated
with the NRAO VLA Sky Survey (NVSS) data at 1.4 GHz to search for
diffuse, central radio emission not associated with radio galaxies or
other non-central diffuse emission. The most important points of our
analysis can be summarized as follows:

\begin{itemize}
\item We iteratively remove all compact sources from the maps, so as
  to extract the central diffuse emission with a wide range of
  morphologies and scales. We attempt to minimize the contribution
  from radio relics and mini-halos, although some contamination cannot
  be ruled out.
\item We employ two independent methods for radio flux extraction,
  based on a average model fit and a direct integration. The flux
  extraction is carried out within a radius $0.5\times r_{500}$, and
  we account for the missing flux outside this region using a common
  stacked radial profile (our results do not depend on this flux
  extrapolation). Due to the large uncertainties, the majority of the
  individual signals are consistent with zero at the $3\sigma$ level.
\item We model the relation between radio power and mass observables
  with a power law, accounting for intrinsic scatter in the
  measurements, uncertainties in the dependent and independent
  variables, and a dropout fraction quantifying the fraction of
  objects not hosting a central, diffuse radio emission component.
\item We run an extensive set of simulations to determine any biases
  that might occur from the filtering method, residual flux from
  bright sources, a point source population below the confusion limit,
  or the regression analysis.
\end{itemize}

\noindent We summarize the main conclusions of our work:

\begin{enumerate}

\item The SZ and X-ray selected cluster samples both show the presence
  of a radio halo population, whose individual and averaged flux
  measurements are generally consistent with previously published
  results. The scaling of the radio halo power with the total cluster
  mass from these two samples are consistent. The intrinsic scatter is
  found to be large.

\item The SZ selected samples based on the Planck PSZ catalog yield a
  low radio halo dropout fraction (i.e. clusters hosting no radio
  halos). For a sub-sample built from the redshift-dependent mass
  limit similar to the Planck cluster cosmology sample, the dropout
  fraction is {\referee roughly $30 \pm 10\%$}. For a constant
  mass-cut PSZ sub-sample with masses $M_{500} > 8\times 10^{14}$
  M$_{\odot}$, the dropout fraction is found to be consistent with
  zero at approximately $15\pm 10$\%, suggesting a nearly
  complete occurrence rate at the very high-mass end.

\item The situation is different in the case of X-ray selection. Using
  a complete sample based on the REFLEX, BCS/eBCS, NORAS and MACS
  cluster catalogs and the same redshift-dependent mass limit, the
  dropout fraction in X-ray selected clusters is roughly $60\pm
  10$\%. The fraction is {\referee effectively unchanged in the constant
    mass selection}. These numbers are fully
  consistent with the general view that radio halos are rare objects,
  with roughly 70\% of the high X-ray luminosity clusters
  ($L_{\mathrm{X}} > 5 \times 10^{44}$ erg/s) being in the radio
  ``off-state".

\item The difference between the SZ and X-ray selections is likely a
  combination of two effects: dissimilar scatter in these two mass
  observables during cluster mergers, and a cool-core bias in X-ray
  flux limited samples. The first argument can be used to explain the
  absence of radio halos in some {\referee early} mergers, and to
  predict a large number of radio halos in clusters that are X-ray
  under-luminous in the late merger phase.
\end{enumerate}

\section*{Acknowledgments}

We would like to thank the anonymous referee for a careful reading of
the manuscript and for many useful suggestions.  We acknowledge the
help of Gregory B. Poole in providing the simulation results of Poole
et al. (2007), and Nabila Aghanim in dealing with the early-release
version of the Planck PSZ catalog.  We thank Arif Babul, Melanie
Johnston-Hollitt, Florian Pacaud, Thomas Reiprich and Nirupam Roy for
useful discussions. MWS acknowledges partial support for this work
from Transregio Programme TR33 of the German Research Foundation
(Deutsche Forschungsgemeinschaft).


\bibliographystyle{mn2e}   
\bibliography{martin}{}    

\begin{thebibliography}{68}
\expandafter\ifx\csname natexlab\endcsname\relax\def\natexlab#1{#1}\fi

\bibitem[{{Arnaud} {et~al}\mbox{.}(2010){Arnaud}, {Pratt}, {Piffaretti},
  {B{\"o}hringer}, {Croston}, \& {Pointecouteau}}]{2010A&A...517A..92A}
{Arnaud} M., {Pratt} G.~W., {Piffaretti} R., {B{\"o}hringer} H., {Croston}
  J.~H., {Pointecouteau} E., 2010, \aap, 517, A92

\bibitem[{{Basu}(2012)}]{2012MNRAS.421L.112B}
{Basu} K., 2012, \mnras, 421, L112

\bibitem[{{Becker}, {White} \& {Helfand}(1995){Becker}, {White}, \&
  {Helfand}}]{1995ApJ...450..559B}
{Becker} R.~H., {White} R.~L., {Helfand} D.~J., 1995, \apj, 450, 559

\bibitem[{{Blasi} \& {Colafrancesco}(1999)}]{1999APh....12..169B}
{Blasi} P., {Colafrancesco} S., 1999, Astroparticle Physics, 12, 169

\bibitem[{{B{\"o}hringer} {et~al}\mbox{.}(2004){B{\"o}hringer}, {Schuecker},
  {Guzzo}, {Collins}, {Voges}, {Cruddace}, {Ortiz-Gil}, {Chincarini}, {De
  Grandi}, {Edge}, {MacGillivray}, {Neumann}, {Schindler}, \&
  {Shaver}}]{2004A&A...425..367B}
{B{\"o}hringer} H. {et~al.}, 2004, \aap, 425, 367

\bibitem[{{B{\"o}hringer} {et~al}\mbox{.}(2000){B{\"o}hringer}, {Voges},
  {Huchra}, {McLean}, {Giacconi}, {Rosati}, {Burg}, {Mader}, {Schuecker},
  {Simi{\c c}}, {Komossa}, {Reiprich}, {Retzlaff}, \&
  {Tr{\"u}mper}}]{2000ApJS..129..435B}
{B{\"o}hringer} H. {et~al.}, 2000, \apjs, 129, 435

\bibitem[{{Bonafede} {et~al}\mbox{.}(2009){Bonafede}, {Giovannini}, {Feretti},
  {Govoni}, \& {Murgia}}]{2009A&A...494..429B}
{Bonafede} A., {Giovannini} G., {Feretti} L., {Govoni} F., {Murgia} M., 2009,
  \aap, 494, 429

\bibitem[{{Brown} {et~al}\mbox{.}(2011){Brown}, {Emerick}, {Rudnick}, \&
  {Brunetti}}]{2011ApJ...740L..28B}
{Brown} S., {Emerick} A., {Rudnick} L., {Brunetti} G., 2011, \apjl, 740, L28

\bibitem[{{Brunetti} {et~al}\mbox{.}(2009){Brunetti}, {Cassano}, {Dolag}, \&
  {Setti}}]{2009A&A...507..661B}
{Brunetti} G., {Cassano} R., {Dolag} K., {Setti} G., 2009, \aap, 507, 661

\bibitem[{{Brunetti} {et~al}\mbox{.}(2001){Brunetti}, {Setti}, {Feretti}, \&
  {Giovannini}}]{2001MNRAS.320..365B}
{Brunetti} G., {Setti} G., {Feretti} L., {Giovannini} G., 2001, \mnras, 320,
  365

\bibitem[{{Buote}(2001)}]{2001ApJ...553L..15B}
{Buote} D.~A., 2001, \apjl, 553, L15

\bibitem[{{Cassano} {et~al}\mbox{.}(2012){Cassano}, {Brunetti}, {Norris},
  {R{\"o}ttgering}, {Johnston-Hollitt}, \& {Trasatti}}]{2012A&A...548A.100C}
{Cassano} R., {Brunetti} G., {Norris} R.~P., {R{\"o}ttgering} H.~J.~A.,
  {Johnston-Hollitt} M., {Trasatti} M., 2012, \aap, 548, A100

\bibitem[{{Cassano} {et~al}\mbox{.}(2010){Cassano}, {Brunetti},
  {R{\"o}ttgering}, \& {Br{\"u}ggen}}]{2010A&A...509A..68C}
{Cassano} R., {Brunetti} G., {R{\"o}ttgering} H.~J.~A., {Br{\"u}ggen} M., 2010,
  \aap, 509, A68

\bibitem[{{Cassano} {et~al}\mbox{.}(2007){Cassano}, {Brunetti}, {Setti},
  {Govoni}, \& {Dolag}}]{2007MNRAS.378.1565C}
{Cassano} R., {Brunetti} G., {Setti} G., {Govoni} F., {Dolag} K., 2007, \mnras,
  378, 1565

\bibitem[{{Cassano} {et~al}\mbox{.}(2013){Cassano}, {Ettori}, {Brunetti},
  {Giacintucci}, {Pratt}, {Venturi}, {Kale}, {Dolag}, \&
  {Markevitch}}]{2013ApJ...777..141C}
{Cassano} R. {et~al.}, 2013, \apj, 777, 141

\bibitem[{{Condon} {et~al}\mbox{.}(1998){Condon}, {Cotton}, {Greisen}, {Yin},
  {Perley}, {Taylor}, \& {Broderick}}]{1998AJ....115.1693C}
{Condon} J.~J., {Cotton} W.~D., {Greisen} E.~W., {Yin} Q.~F., {Perley} R.~A.,
  {Taylor} G.~B., {Broderick} J.~J., 1998, \aj, 115, 1693

\bibitem[{{Coppin} {et~al}\mbox{.}(2011){Coppin}, {Geach}, {Smail}, {Dunne},
  {Edge}, {Ivison}, {Maddox}, {Auld}, {Baes}, {Buttiglione}, {Cava},
  {Clements}, {Cooray}, {Dariush}, {de Zotti}, {Dye}, {Eales}, {Fritz},
  {Hopwood}, {Ibar}, {Jarvis}, {Micha{\l}owski}, {Murphy}, {Negrello},
  {Pascale}, {Pohlen}, {Rigby}, {Rodighiero}, {Scott}, {Serjeant}, {Smith},
  {Temi}, \& {van der Werf}}]{2011MNRAS.416..680C}
{Coppin} K.~E.~K. {et~al.}, 2011, \mnras, 416, 680

\bibitem[{{Dennison}(1980)}]{1980ApJ...239L..93D}
{Dennison} B., 1980, \apjl, 239, L93

\bibitem[{{Donnert} {et~al}\mbox{.}(2013){Donnert}, {Dolag}, {Brunetti}, \&
  {Cassano}}]{2013MNRAS.429.3564D}
{Donnert} J., {Dolag} K., {Brunetti} G., {Cassano} R., 2013, \mnras, 429, 3564

\bibitem[{{Ebeling} {et~al}\mbox{.}(2000){Ebeling}, {Edge}, {Allen},
  {Crawford}, {Fabian}, \& {Huchra}}]{2000MNRAS.318..333E}
{Ebeling} H., {Edge} A.~C., {Allen} S.~W., {Crawford} C.~S., {Fabian} A.~C.,
  {Huchra} J.~P., 2000, \mnras, 318, 333

\bibitem[{{Ebeling} {et~al}\mbox{.}(1998){Ebeling}, {Edge}, {Bohringer},
  {Allen}, {Crawford}, {Fabian}, {Voges}, \& {Huchra}}]{1998MNRAS.301..881E}
{Ebeling} H., {Edge} A.~C., {Bohringer} H., {Allen} S.~W., {Crawford} C.~S.,
  {Fabian} A.~C., {Voges} W., {Huchra} J.~P., 1998, \mnras, 301, 881

\bibitem[{{Ebeling}, {Edge} \& {Henry}(2001){Ebeling}, {Edge}, \&
  {Henry}}]{2001ApJ...553..668E}
{Ebeling} H., {Edge} A.~C., {Henry} J.~P., 2001, \apj, 553, 668

\bibitem[{{Eckert}, {Molendi} \& {Paltani}(2011){Eckert}, {Molendi}, \&
  {Paltani}}]{2011A&A...526A..79E}
{Eckert} D., {Molendi} S., {Paltani} S., 2011, \aap, 526, A79

\bibitem[{{En{\ss}lin} {et~al}\mbox{.}(2011){En{\ss}lin}, {Pfrommer},
  {Miniati}, \& {Subramanian}}]{2011A&A...527A..99E}
{En{\ss}lin} T., {Pfrommer} C., {Miniati} F., {Subramanian} K., 2011, \aap,
  527, A99

\bibitem[{{Ensslin} {et~al}\mbox{.}(1998){Ensslin}, {Biermann}, {Klein}, \&
  {Kohle}}]{1998A&A...332..395E}
{Ensslin} T.~A., {Biermann} P.~L., {Klein} U., {Kohle} S., 1998, \aap, 332, 395

\bibitem[{{Feretti} {et~al}\mbox{.}(2012){Feretti}, {Giovannini}, {Govoni}, \&
  {Murgia}}]{2012A&ARv..20...54F}
{Feretti} L., {Giovannini} G., {Govoni} F., {Murgia} M., 2012, \aapr, 20, 54

\bibitem[{{Giovannini} {et~al}\mbox{.}(2009){Giovannini}, {Bonafede},
  {Feretti}, {Govoni}, {Murgia}, {Ferrari}, \& {Monti}}]{2009A&A...507.1257G}
{Giovannini} G., {Bonafede} A., {Feretti} L., {Govoni} F., {Murgia} M.,
  {Ferrari} F., {Monti} G., 2009, \aap, 507, 1257

\bibitem[{{Giovannini} {et~al}\mbox{.}(2011){Giovannini}, {Feretti}, {Girardi},
  {Govoni}, {Murgia}, {Vacca}, \& {Bagchi}}]{2011A&A...530L...5G}
{Giovannini} G., {Feretti} L., {Girardi} M., {Govoni} F., {Murgia} M., {Vacca}
  V., {Bagchi} J., 2011, \aap, 530, L5

\bibitem[{{Govoni} {et~al}\mbox{.}(2001){Govoni}, {En{\ss}lin}, {Feretti}, \&
  {Giovannini}}]{2001A&A...369..441G}
{Govoni} F., {En{\ss}lin} T.~A., {Feretti} L., {Giovannini} G., 2001, \aap,
  369, 441

\bibitem[{{Govoni} {et~al}\mbox{.}(2004){Govoni}, {Markevitch}, {Vikhlinin},
  {van Speybroeck}, {Feretti}, \& {Giovannini}}]{2004ApJ...605..695G}
{Govoni} F., {Markevitch} M., {Vikhlinin} A., {van Speybroeck} L., {Feretti}
  L., {Giovannini} G., 2004, \apj, 605, 695

\bibitem[{{Hoeft} \& {Br{\"u}ggen}(2007)}]{2007MNRAS.375...77H}
{Hoeft} M., {Br{\"u}ggen} M., 2007, \mnras, 375, 77

\bibitem[{{Hoeft} {et~al}\mbox{.}(2011){Hoeft}, {Nuza}, {Gottl{\"o}ber}, {van
  Weeren}, {R{\"o}ttgering}, \& {Br{\"u}ggen}}]{2011JApA...32..509H}
{Hoeft} M., {Nuza} S.~E., {Gottl{\"o}ber} S., {van Weeren} R.~J.,
  {R{\"o}ttgering} H.~J.~A., {Br{\"u}ggen} M., 2011, Journal of Astrophysics
  and Astronomy, 32, 509

\bibitem[{{Hogg}, {Bovy} \& {Lang}(2010){Hogg}, {Bovy}, \&
  {Lang}}]{2010arXiv1008.4686H}
{Hogg} D.~W., {Bovy} J., {Lang} D., 2010, ArXiv e-prints

\bibitem[{{Kale} {et~al}\mbox{.}(2013){Kale}, {Venturi}, {Giacintucci},
  {Dallacasa}, {Cassano}, {Brunetti}, {Macario}, \&
  {Athreya}}]{2013A&A...557A..99K}
{Kale} R., {Venturi} T., {Giacintucci} S., {Dallacasa} D., {Cassano} R.,
  {Brunetti} G., {Macario} G., {Athreya} R., 2013, \aap, 557, A99

\bibitem[{{Kelly}(2007)}]{2007ApJ...665.1489K}
{Kelly} B.~C., 2007, \apj, 665, 1489

\bibitem[{{Kushnir}, {Katz} \& {Waxman}(2009){Kushnir}, {Katz}, \&
  {Waxman}}]{2009JCAP...09..024K}
{Kushnir} D., {Katz} B., {Waxman} E., 2009, \jcap, 9, 24

\bibitem[{{Liang} {et~al}\mbox{.}(2000){Liang}, {Hunstead}, {Birkinshaw}, \&
  {Andreani}}]{2000ApJ...544..686L}
{Liang} H., {Hunstead} R.~W., {Birkinshaw} M., {Andreani} P., 2000, \apj, 544,
  686

\bibitem[{{Lin} \& {Mohr}(2004)}]{2004ApJ...617..879L}
{Lin} Y., {Mohr} J.~J., 2004, \apj, 617, 879

\bibitem[{{Mittal} {et~al}\mbox{.}(2009){Mittal}, {Hudson}, {Reiprich}, \&
  {Clarke}}]{2009A&A...501..835M}
{Mittal} R., {Hudson} D.~S., {Reiprich} T.~H., {Clarke} T., 2009, \aap, 501,
  835

\bibitem[{{Norris} {et~al}\mbox{.}(2011){Norris}, {Hopkins}, {Afonso}, {Brown},
  {Condon}, {Dunne}, {Feain}, {Hollow}, {Jarvis}, {Johnston-Hollitt}, {Lenc},
  {Middelberg}, {Padovani}, {Prandoni}, {Rudnick}, {Seymour}, {Umana},
  {Andernach}, {Alexander}, {Appleton}, {Bacon}, {Banfield}, {Becker}, {Brown},
  {Ciliegi}, {Jackson}, {Eales}, {Edge}, {Gaensler}, {Giovannini}, {Hales},
  {Hancock}, {Huynh}, {Ibar}, {Ivison}, {Kennicutt}, {Kimball}, {Koekemoer},
  {Koribalski}, {L{\'o}pez-S{\'a}nchez}, {Mao}, {Murphy}, {Messias},
  {Pimbblet}, {Raccanelli}, {Randall}, {Reiprich}, {Roseboom},
  {R{\"o}ttgering}, {Saikia}, {Sharp}, {Slee}, {Smail}, {Thompson}, {Urquhart},
  {Wall}, \& {Zhao}}]{2011PASA...28..215N}
{Norris} R.~P. {et~al.}, 2011, \pasa, 28, 215

\bibitem[{{Nuza} {et~al}\mbox{.}(2012){Nuza}, {Hoeft}, {van Weeren},
  {Gottl{\"o}ber}, \& {Yepes}}]{2012MNRAS.420.2006N}
{Nuza} S.~E., {Hoeft} M., {van Weeren} R.~J., {Gottl{\"o}ber} S., {Yepes} G.,
  2012, \mnras, 420, 2006

\bibitem[{{Petrosian}(2001)}]{2001ApJ...557..560P}
{Petrosian} V., 2001, \apj, 557, 560

\bibitem[{{Piffaretti} {et~al}\mbox{.}(2011){Piffaretti}, {Arnaud}, {Pratt},
  {Pointecouteau}, \& {Melin}}]{2011A&A...534A.109P}
{Piffaretti} R., {Arnaud} M., {Pratt} G.~W., {Pointecouteau} E., {Melin} J.-B.,
  2011, \aap, 534, A109

\bibitem[{{Planck Collaboration}(2011{\natexlab{a}})}]{2011A&A...536A...8P}
{Planck Collaboration}, 2011{\natexlab{a}}, \aap, 536, A8

\bibitem[{{Planck Collaboration}(2011{\natexlab{b}})}]{2011A&A...536A..11P}
{Planck Collaboration}, 2011{\natexlab{b}}, \aap, 536, A11

\bibitem[{{Planck Collaboration}(2013{\natexlab{a}})}]{2013arXiv1303.5080P}
{Planck Collaboration}, 2013{\natexlab{a}}, ArXiv:astro-ph/1303.5080

\bibitem[{{Planck Collaboration}(2013{\natexlab{b}})}]{2013arXiv1303.5089P}
{Planck Collaboration}, 2013{\natexlab{b}}, ArXiv:astro-ph/1303.5089

\bibitem[{{Poole} {et~al}\mbox{.}(2007){Poole}, {Babul}, {McCarthy}, {Fardal},
  {Bildfell}, {Quinn}, \& {Mahdavi}}]{2007MNRAS.380..437P}
{Poole} G.~B., {Babul} A., {McCarthy} I.~G., {Fardal} M.~A., {Bildfell} C.~J.,
  {Quinn} T., {Mahdavi} A., 2007, \mnras, 380, 437

\bibitem[{{R{\"o}ttgering} {et~al}\mbox{.}(2011){R{\"o}ttgering}, {Afonso},
  {Barthel}, {Batejat}, {Best}, {Bonafede}, {Br{\"u}ggen}, {Brunetti},
  {Chy{\.z}y}, {Conway}, {Gasperin}, {Ferrari}, {Haverkorn}, {Heald}, {Hoeft},
  {Jackson}, {Jarvis}, {Ker}, {Lehnert}, {Macario}, {McKean}, {Miley},
  {Morganti}, {Oosterloo}, {Orr{\`u}}, {Pizzo}, {Rafferty}, {Shulevski},
  {Tasse}, {Bemmel}, {van der Tol}, {van Weeren}, {Verheijen}, {White}, \&
  {Wise}}]{2011JApA...32..557R}
{R{\"o}ttgering} H. {et~al.}, 2011, Journal of Astrophysics and Astronomy, 32,
  557

\bibitem[{{Rudnick}(2002)}]{2002PASP..114..427R}
{Rudnick} L., 2002, \pasp, 114, 427

\bibitem[{{Rudnick} \& {Lemmerman}(2009)}]{2009ApJ...697.1341R}
{Rudnick} L., {Lemmerman} J.~A., 2009, \apj, 697, 1341

\bibitem[{{Russell} {et~al}\mbox{.}(2011){Russell}, {van Weeren}, {Edge},
  {McNamara}, {Sanders}, {Fabian}, {Baum}, {Canning}, {Donahue}, \&
  {O'Dea}}]{2011MNRAS.417L...1R}
{Russell} H.~R. {et~al.}, 2011, \mnras, 417, L1

\bibitem[{{Sayers} {et~al}\mbox{.}(2013){Sayers}, {Mroczkowski}, {Czakon},
  {Golwala}, {Mantz}, {Ameglio}, {Downes}, {Koch}, {Lin}, {Molnar},
  {Moustakas}, {Muchovej}, {Pierpaoli}, {Shitanishi}, {Siegel}, \&
  {Umetsu}}]{2013ApJ...764..152S}
{Sayers} J. {et~al.}, 2013, \apj, 764, 152

\bibitem[{{Schlickeiser}, {Sievers} \& {Thiemann}(1987){Schlickeiser},
  {Sievers}, \& {Thiemann}}]{1987A&A...182...21S}
{Schlickeiser} R., {Sievers} A., {Thiemann} H., 1987, \aap, 182, 21

\bibitem[{{Schuecker} {et~al}\mbox{.}(2001){Schuecker}, {B{\"o}hringer},
  {Reiprich}, \& {Feretti}}]{2001A&A...378..408S}
{Schuecker} P., {B{\"o}hringer} H., {Reiprich} T.~H., {Feretti} L., 2001, \aap,
  378, 408

\bibitem[{{Sehgal} {et~al}\mbox{.}(2010){Sehgal}, {Bode}, {Das},
  {Hernandez-Monteagudo}, {Huffenberger}, {Lin}, {Ostriker}, \&
  {Trac}}]{2010ApJ...709..920S}
{Sehgal} N., {Bode} P., {Das} S., {Hernandez-Monteagudo} C., {Huffenberger} K.,
  {Lin} Y.-T., {Ostriker} J.~P., {Trac} H., 2010, \apj, 709, 920

\bibitem[{{Skillman} {et~al}\mbox{.}(2011){Skillman}, {Hallman}, {O'Shea},
  {Burns}, {Smith}, \& {Turk}}]{2011ApJ...735...96S}
{Skillman} S.~W., {Hallman} E.~J., {O'Shea} B.~W., {Burns} J.~O., {Smith}
  B.~D., {Turk} M.~J., 2011, \apj, 735, 96

\bibitem[{{Sommer} {et~al}\mbox{.}(2011){Sommer}, {Basu}, {Pacaud}, {Bertoldi},
  \& {Andernach}}]{2011A&A...529A.124S}
{Sommer} M.~W., {Basu} K., {Pacaud} F., {Bertoldi} F., {Andernach} H., 2011,
  \aap, 529, A124

\bibitem[{{Sun}(2009)}]{2009ApJ...704.1586S}
{Sun} M., 2009, \apj, 704, 1586

\bibitem[{{Sunyaev} \& {Zeldovich}(1980)}]{1980ARA&A..18..537S}
{Sunyaev} R.~A., {Zeldovich} I.~B., 1980, \araa, 18, 537

\bibitem[{{Sunyaev} \& {Zeldovich}(1972)}]{1972A&A....20..189S}
{Sunyaev} R.~A., {Zeldovich} Y.~B., 1972, \aap, 20, 189

\bibitem[{{Sutter} \& {Ricker}(2012)}]{2012ApJ...759...92S}
{Sutter} P.~M., {Ricker} P.~M., 2012, \apj, 759, 92

\bibitem[{{Tinker} {et~al}\mbox{.}(2008){Tinker}, {Kravtsov}, {Klypin},
  {Abazajian}, {Warren}, {Yepes}, {Gottl{\"o}ber}, \&
  {Holz}}]{2008ApJ...688..709T}
{Tinker} J., {Kravtsov} A.~V., {Klypin} A., {Abazajian} K., {Warren} M.,
  {Yepes} G., {Gottl{\"o}ber} S., {Holz} D.~E., 2008, \apj, 688, 709

\bibitem[{{van Weeren} {et~al}\mbox{.}(2011{\natexlab{a}}){van Weeren},
  {Br{\"u}ggen}, {R{\"o}ttgering}, \& {Hoeft}}]{2011MNRAS.418..230V}
{van Weeren} R.~J., {Br{\"u}ggen} M., {R{\"o}ttgering} H.~J.~A., {Hoeft} M.,
  2011{\natexlab{a}}, \mnras, 418, 230

\bibitem[{{van Weeren} {et~al}\mbox{.}(2011{\natexlab{b}}){van Weeren},
  {Br{\"u}ggen}, {R{\"o}ttgering}, {Hoeft}, {Nuza}, \&
  {Intema}}]{2011A&A...533A..35V}
{van Weeren} R.~J., {Br{\"u}ggen} M., {R{\"o}ttgering} H.~J.~A., {Hoeft} M.,
  {Nuza} S.~E., {Intema} H.~T., 2011{\natexlab{b}}, \aap, 533, A35

\bibitem[{{Venturi} {et~al}\mbox{.}(2008){Venturi}, {Giacintucci}, {Dallacasa},
  {Cassano}, {Brunetti}, {Bardelli}, \& {Setti}}]{2008A&A...484..327V}
{Venturi} T., {Giacintucci} S., {Dallacasa} D., {Cassano} R., {Brunetti} G.,
  {Bardelli} S., {Setti} G., 2008, \aap, 484, 327

\bibitem[{{Wen} \& {Han}(2013)}]{2013MNRAS.436..275W}
{Wen} Z.~L., {Han} J.~L., 2013, \mnras, 436, 275

\bibitem[{{White} {et~al}\mbox{.}(1997){White}, {Becker}, {Helfand}, \&
  {Gregg}}]{1997ApJ...475..479W}
{White} R.~L., {Becker} R.~H., {Helfand} D.~J., {Gregg} M.~D., 1997, \apj, 475,
  479

\end{thebibliography}

\label{lastpage}

\end{document}